\documentclass{emulateapj}
\usepackage{epsfig}
\usepackage{natbib}
\usepackage{amsmath}

\newlength{\colwidth}

\newcommand{\be}{\begin{equation}}
\newcommand{\ee}{\end{equation}}
\newcommand{\bee}{\begin{eqnarray}}
\newcommand{\eee}{\end{eqnarray}}

\newcommand{\msol}{\hbox{${\rm M}_\odot$}}

\newcommand{\kms}{km s$^{-1}$}

\newcommand{\lsim}{\lower.5ex\hbox{\ltsima}}
\newcommand{\gsim}{\lower.5ex\hbox{\gtsima}}
\newcommand{\ltsima}{$\; \buildrel < \over \sim \;$}
\newcommand{\gtsima}{$\; \buildrel > \over \sim \;$}

\usepackage[usenames]{color}

\definecolor{purple}{RGB}{160,32,240}

\newcommand{\ie}{\emph{i.e.}}
\newcommand{\eg}{\emph{e.g.}}

\shortauthors{LIU ET AL.}
\shorttitle{How Common are the Magellanic Clouds? }
\begin{document}

\title{How Common are the Magellanic Clouds? }
\author{Lulu Liu, Brian F. Gerke, Risa H. Wechsler, Peter S. Behroozi,
  and Michael T. Busha\altaffilmark{1}
}
\affil{Kavli Institute for Particle Astrophysics and Cosmology \\
452 Lomita Mall, Stanford University, Stanford, CA, 94305\\
SLAC National Accelerator Laboratory, 2575 Sand Hill Rd., MS 29, Menlo Park, CA, 94025\\
{\tt lululiu, bgerke, rwechsler, behroozi, busha@slac.stanford.edu}}

\altaffiltext{1}{present address: Institute for Theoretical Physics, Unviersity of Zurich, 8057 Zurich, Switzerland} 

\begin{abstract}
  We introduce a probabilistic approach to the problem of counting
  dwarf satellites around host galaxies in databases with limited
  redshift information.  This technique is used to investigate the
  occurrence of satellites with luminosities similar to the Magellanic
  Clouds around hosts with properties similar to the Milky Way in the
  object catalog of the Sloan Digital Sky Survey.  Our analysis uses
  data from SDSS Data Release 7, selecting candidate Milky-Way-like
  hosts from the spectroscopic catalog and candidate analogs of the
  Magellanic Clouds from the photometric catalog. Our principal result
  is the probability for a Milky-Way-like galaxy to host $N_{sat}$
  close satellites with luminosities similar to the Magellanic Clouds.
  We find that 81 percent of galaxies like the Milky Way have no
  such satellites within a radius of 150 kpc, 11 percent have one, and
  only 3.5 percent of hosts have two.  The probabilities are robust 
  to changes in host and satellite selection criteria,
  background-estimation technique, and survey depth.  These results
  demonstrate that the Milky Way has significantly more satellites
  than a typical galaxy of its luminosity; this fact is useful for
  understanding the larger cosmological context of our home galaxy.
\end{abstract}


\section{Introduction}

Our home galaxy, the Milky Way, is in many respects the best studied
galaxy in the Universe.  There are numerous measurements that can only
be made in the Milky Way, including detailed studies of resolved
stellar populations and the detection and dynamical measurements of
the faintest satellite galaxies.  Furthermore, the Milky Way is a
critical testbed for dark matter studies, as it is one of the only
places where self-annihilation or weak interactions can be directly
observed.  As such, studies of the Milky Way have long provided key
insights into aspects of cosmology and galaxy formation.

To fully interpret this panoply of observations of the Milky Way (MW)
in the context of a cosmological model, it is essential to understand
whether or not the MW is a typical galaxy of its mass or
luminosity. One of the most cosmologically interesting statistical
properties of the MW that can be readily studied is the number and
properties of its satellites.  It has been apparent for more than a
decade that N-body simulations of Galaxy-sized (\ie, M $\sim
10^{12}M_\odot$) dark-matter halos predict an abundance of low-mass
subhalos that exceeds the observed population of MW dwarf satellites
by more than an order of magnitude (\citealt{Moore99b, Klypin99}; for
recent reviews, see \citealt{bullock10, kravtsov10}); this is the
so-called ``missing satellites problem.''  A number of different
theoretical solutions to this problem have been proposed, focusing
either on reducing the small-scale power in the dark-matter power
spectrum, or on suppressing galaxy formation in low-mass halos
\citep[see, \eg,][and references therein]{Madau08, Busha10a}.

In the last several years, it has become apparent that some of this
discrepancy was due to galaxies that had not yet been observed.  The
unprecedented deep, wide-field imaging from the Sloan Digital Sky
Survey \citep[SDSS;][]{york_etal:00} has yielded detections of a
substantial number of previously unknown dwarf companions to the MW
\citep[\eg,][]{Belokurov07b, Walsh09}.  This has led to a reassessment of the
missing satellites problem from the observational side, with the
result that proper accounting for the detectability of MW dwarf
satellites \citep{Koposov08} results in substantial upward corrections
to the satellite luminosity function.  This leads to the prediction
that the MW hosts hundreds of faint satellites that are currently
undetected \citep{Tollerud08, Walsh09}.  

By contrast, there has been some indication that the
missing-satellites problem might reverse itself at high masses:
high-resolution Galactic-halo simulations generally have too
\emph{few} subhalos in the mass range of the Large and Small
Magellanic Clouds (LMC and SMC) \citep[\eg,][]{Madau08}.  However, it
has been difficult to draw robust conclusions on this point, since the
numbers of high-mass subhalos, such as those that might host the
Magellanic Clouds, are few in number for any individual Galactic-halo
simulation, and in any case the abundance of such massive subhalos
might be suppressed by the limited number of long-wavelength
density-fluctuation modes in a small simulation volume.

With the completion of recent high-resolution N-body simulations over
cosmological volumes, such as the Millennium II and Bolshoi
simulations \citep{MillenniumII, Bolshoi}, it has become possible to
probe analogs of the MW-LMC-SMC system with greatly improved
statistical significance, since these simulations can resolve LMC- and
SMC-mass subhalos in large numbers of MW-mass halos.  For example,
\citet{MBK10} found that MW-mass halos (with $M_{halo} \sim 10^{12}
M_\odot$) seldom host subhalos that can be identified as analogs to
the Magellanic Clouds, with less than $10\%$ of MW-sized halos hosting
two such subhalos.  Similar results are seen in the Bolshoi
simulation, as will be discussed in a companion paper to this one
\citep{Busha10b}.  It thus appears that the MW-LMC-SMC system is
somewhat atypical in the context of the Cold Dark Matter paradigm.

On its face, this theoretical result seems to be mildly
anti-Copernican, so it is especially important to confront it with
observational data.  But any satellite-counting exercise is
classically complicated by the faintness of the satellites, which
makes obtaining redshift information difficult.  Redshift-space
studies of satellite dynamics that have aimed to probe galaxy halo
masses \citep[\eg,][]{zaritsky93, prada_etal:03, conroy05,
  conroy07a} or profiles \citep[\eg][]{chen_etal:06} have typically
had redshift information for $\la 1$ 
satellite per host galaxy, even while using criteria for satellite
selection that are significantly more relaxed than we would like to
use to select LMC/SMC analogs.  Indeed, as discussed in
Section~\ref{sec:spectrosat} below, even the vast SDSS spectroscopic
database contains less than $100$ MW-like systems that host one or
more MC-like satellites with redshifts.

A common method for overcoming the lack of spectroscopic information
for satellites has been to count candidate satellites around bright
hosts in photometric data and to statistically subtract or correct for
the contribution of foreground and background objects (hereafter, we
will simply use the term ``background'' as a shorthand to refer to both
foreground and background objects).  In a
pioneering paper, \citet{holmberg69} found that nearby bright spiral
galaxies, with a rather wide range of luminosities, typically hosted
between 0 and 5 satellites brighter than an absolute magnitude around
$-10.6$.   \citet{lorrimer94} carried out a similar analysis and found
that galaxies brighter than $M_{B_T} = -18.5$ host 1.1 satellites in
the range $-16\ge M_{B_T} \ge -18$, on average, with fewer satellites
around spirals (0.5 on average) than ellipticals (1.8).  Both of those
studies accounted for background contamination by
subtracting off the average number of faint galaxies in nearby fields
from the counts around bright galaxies; therefore, they were only able to
measure the average number of satellites around bright galaxies; as
discussed in Section~\ref{sec:methods}, they cannot address the
\emph{probability} of hosting a certain number of satellites, which is
what we have set out to do here.   

Recently, \citet{james10} carried out a
quasi-spectroscopic analysis by targeting 143 bright galaxies of known
redshift for follow-up with narrow-band imaging centered near the
expected wavelength of H$\alpha$ at the redshift of the target galaxy.
This allowed them to count the number of roughly MC-like star-forming
objects within a few hundred \kms of each host, yielding a plausible
measurement of the satellite number-count distribution.  In broad
agreement with simulations, they find that roughly two-thirds of their
target galaxies have zero such satellites, while only $\sim 5\%$ have
two.  This clearly confirms that the Magellanic Clouds are indeed
quite rare.  There is significant uncertainty in the details, however,
owing both to the small sample size and to the width of the imaging
band, which will detect galaxies in H$\alpha$ up to 30 Mpc away from
the host along the line of sight, so the potential for significant
background contamination remains\footnote{Indeed, as we
  will discuss, even in the case of perfect spectroscopic information,
  redshift-space distortions lead to significant contamination from
  interlopers, which must be accounted for.}.  In addition, comparison
to the subhalo population in simulations is difficult, since only
star-forming galaxies are selected.

In this paper, we employ the enormous statistical power of the SDSS to
obtain a statistically robust result for the frequency of LMC and SMC
analogs in galaxies like the MW.  We use the main SDSS spectroscopic
catalog to identify a sample of $>2\times 10^5$ isolated galaxies with
luminosities similar to the MW.  As in the studies above, we then
count photometric companions around host galaxies with known
redshifts, but here we introduce a new technique for statistical
background removal that allows us to recover the true probability
distribution of satellite number counts around these hosts,
$P(N_{sat})$.  To do this, we make use of the fact that our measured
number counts represent a convolution of the true satellite
distribution with the distribution of background counts.  We can also
measure the latter distribution in the data, and then a simple
deconvolution yields the desired result.  We pay careful attention to
the details of our background-estimation techniques to ensure that we
account for all possible sources of systematic error, particularly
those arising from the clustering of background galaxies with our
hosts.

Our principal result is that only $3.5\% \pm 1.4\%$ of galaxies with
luminosity similar to the MW host two satellites similar to the
Magellanic Clouds within a radius of 150 kpc.  When we split the sample
into red and blue hosts, we find that excluding red-sequence galaxies
from our sample of hosts has no significant effect on the probability
of hosting any number of bright satellites.  This confirms that the
MW-LMC-SMC system is indeed quite unusual compared to the population
of isolated galaxies with similar luminosity and color.  These results
are also broadly consistent with the predictions from simulations
\citep{MBK10}.  We present a detailed comparison with the predictions
of high-resolution $\Lambda$CDM simulations in a companion paper
\citep{Busha10b}; the general conclusion from both simulations and
observations is that the MW-LMC-SMC system is quite rare.

This paper is structured as follows.  First, we describe the SDSS
dataset and our criteria for selecting analogs to the MW, LMC and SMC
in Section~\ref{sec:data}.  In Section~\ref{sec:spectrosat} we perform
a preliminary analysis on SDSS galaxies with spectroscopically
identified satellites.  Finding it difficult to cleanly interpret
these results, we then move on to develop our photometric background
subtraction method in Section~\ref{sec:methods}, being careful to
fully account for the statistical and systematic error budget.  We
perform the counting and background-correction exercise in two
different ways, which have different approaches to handling systematic
errors.  We obtain similar results for these two different procedures,
which we present in Section~\ref{sec:results}.  In that section,
we also discuss the sensitivity of these results to a number of assumptions
in the analysis.  We also perform an additional analysis in the deep
photometric data from SDSS Stripe 82 to test the robustness of our
analysis.  We discuss the implications of our results in
Section~\ref{sec:conclusions}.

Throughout, distances and absolute magnitudes are calculated using a
flat $\Lambda$CDM cosmological model with $\Omega_m = 0.3$.  All
distances quoted in this paper are given in physical (rather than
comoving) units, and all distances and absolute magnitudes derived
from SDSS data assume a Hubble constant of $H_0 = 70$ km s$^{-1}$
Mpc$^{-1}$.

\section{Data and Sample Selection}
\label{sec:data}

\subsection{SDSS Catalogs}
We use data from the spectroscopic and imaging catalogs of the Sloan
Digital Sky Survey (SDSS) seventh data release
\citep[DR7;][]{Abazajian09}.  Because our potential MW-LMC-SMC analogs
are at very low redshift, and because we require them to be more than
500 kpc from a survey edge (see Section~\ref{sec:MWhosts}), this
limits the area used for the analysis.  The main sample of MW-sized
hosts is selected from among the spectroscopic targets in a
contiguous, 3350 square degree section of the Northern Galactic Cap.
The deeper imaging in Stripe 82 (an approximately 280 square-degree
strip along the celestial equator) allows for a second, much smaller
sample of MW analogs extending to slightly higher redshifts in the
southern sky (we analyze this data separately to check the robustness
of our methods; see Section~\ref{sec:stripe82}).  Spectra and
$k$-corrected luminosity values are taken from the NYU Value Added
Galaxy Catalog (NYU-VAGC) \citep{Blanton05} \footnote{\tt
  http://sdss.physics.nyu.edu/vagc/}.

The $r$-band magnitude limit of the main (non-QSO) spectroscopic
galaxy catalog is 17.77.  We use this limit along with the
$PRIMTARGET$ type designation to isolate only those members of the
main catalog identified as galaxies.  We refer to the resulting
catalog as the spectroscopic galaxy sample.  Since we are interested
in relatively faint satellites (between 2 and 4 magnitudes dimmer than
their hosts), the spectroscopic catalog alone is insufficient for our
purposes.  In order to collect ample statistics we must also use the
SDSS photometric data, which is complete down to at least $r\sim 21.5$ for
extended sources.

Our general strategy is to select candidate MW-sized host galaxies
from the spectroscopic sample and conduct our search for satellites
around these objects within the deeper imaging catalog.  From the
NYU-VAGC we obtain the $k$-corrected absolute magnitudes for potential
hosts computed using spectroscopic redshift information.  From the
imaging catalog we obtain apparent magnitudes of galaxies and
photometric redshift information.  In our core analysis we use
photometric redshift probability distributions, $p(z)$, determined by
\cite{cunha} for the DR7 SDSS imaging catalog using an artificial
neural network algorithm.

\subsection{Selection of Milky Way-Sized Central Galaxies}
\label{sec:MWhosts}

\subsubsection{Luminosity Requirements}
As discussed in Section~\ref{sec:satellites}, we count candidate
LMC/SMC analogs by looking for galaxies 2 to 4 magnitudes fainter than
their hosts.  Thus, we limit our pool of potential hosts to galaxies
more than 4 magnitudes brighter than the limit of our photometric
sample.  Objects dimmer than $r=21$ in the imaging catalog are
particularly prone to catastrophic photo-$z$ failures, owing to their
large photometric errors and the sparseness of the available
spectroscopic training set. We therefore limit the pool of potential
MW-analog hosts to apparent magnitude $r<17$ to avoid this uncertain
regime in the photometric sample (in the deeper co-added stripe 82
data we consider hosts as dim as $r=17.6$).  From this reduced
spectroscopic sample, we select a statistically robust set of
MW-luminosity hosts as follows.

The current best estimate for the absolute magnitude of the MW is $M_V
= -20.9$ in the Vega photometric system \citep{vandenBergh}.  In order
to translate this value to the SDSS photometric system, we convert to
the AB system and also apply an appropriate magnitude correction to
the SDSS $r$ filter (which is the filter that has the strongest
overlap with $V$).  To accomplish the latter conversion, we compute
estimated absolute $^{0.0}V$ and $^{0.1}r$-band
magnitudes\footnote{The superscripts indicate the assumed redshift to
  which the colors $k$-corrected; 0.1 is standard for the VAGC, while
  0.0 is appropriate for the MW.  We assume $h=0.7$ in computing all
  $k$-corrected absolute magnitudes} for a large sample of SDSS
spectroscopic targets, using the {\tt kcorrect} algorithm, version
\texttt{kcorrect v4\_1\_4} \citep{BlantonRoweis}.  We then compute the
mean $V-r$ color for galaxies within $\pm 0.2$ magnitudes of the MW
and apply this average correction to the
measured $M_V$ of the MW.  Because the $V$ and $r$ bands overlap
strongly, this correction is quite small (so, for example, splitting
the sample by color before computing it will not make a significant
difference).  The resulting absolute $^{0.1}r$-band magnitude of the
MW is $M_{0.1,r} = -21.2$.  We consider a galaxy to be a potential
MW-like host if it is within $\pm$0.2 magnitudes of this absolute
magnitude.

It is worth noting in passing that the absolute magnitude of the MW is
difficult to measure and may be subject to quite large uncertainties.
Since these are rather difficult to quantify, we adopt a best-guess
value for the MW luminosity here and defer study of the satellite
population's dependence on host luminosity to future work.

\subsubsection{Isolation Criteria}
\label{sec:iso}

Our aim is to count MC-like satellites around MW-like host galaxies.
To ensure that the satellites we are counting are indeed hosted by 
MW analogs, we require that each candidate host, like the MW itself,
is not itself a satellite of a more massive system.\footnote{Although
  the MW and M31 are gravitationally bound, and M31 may be the more
  massive galaxy, the MW is not classified as a satellite of M31,
  since the two galaxies do not (yet) form a virialized system.} This
criterion is simple to impose if we presume that there is a
monotonic relation between dark-matter halo mass and galaxy
luminosity: we can then impose a radius of isolation ($R_{iso}$),
within which no other similarly luminous galaxy may reside.  More
specifically, a candidate host is eliminated if, within this region,
(1) a galaxy brighter (in absolute magnitude) than $M_{host} + \Delta
M_{iso}$ is found within $\pm 1000$ \kms of the host redshift, or (2)
a galaxy brighter (in apparent magnitude) than $m_{host} + \Delta
M_{iso}$ is found with no redshift information\footnote{This happens
  occasionally, though infrequently, owing to fiber conflicts in the
  SDSS spectrograph.}.
In our primary analysis, we set $\Delta M_{iso} = 0$ and reject only
those candidate hosts with a \emph{brighter} neighbor. In Section
\ref{sec:robust}, we also consider the impact of a more stringent
condition, requiring that our MW analogs have no close neighbors of
similar brightness (up to $\Delta M_{iso} = 2$), and we show that the
results are insensitive to this detail.

The exact choice of $R_{iso}$ naturally involves some trade-offs.  The
primary impact of varying $R_{iso}$ will be to change the purity and
completeness of the sample of isolated hosts.  Here, purity refers to
the fraction of surviving MW analogs which are indeed isolated, that
is, which are not satellites of more massive galaxies.  Completeness
is the fraction of truly isolated hosts that pass through our
isolation filter.  To explore the effect of the $R_{iso}$ parameter on
these statistics, we consider its impact on dark matter halos
identified in a cosmological N-body simulation.

\begin{figure}[t!]
\centering
\includegraphics[width=3.3in]{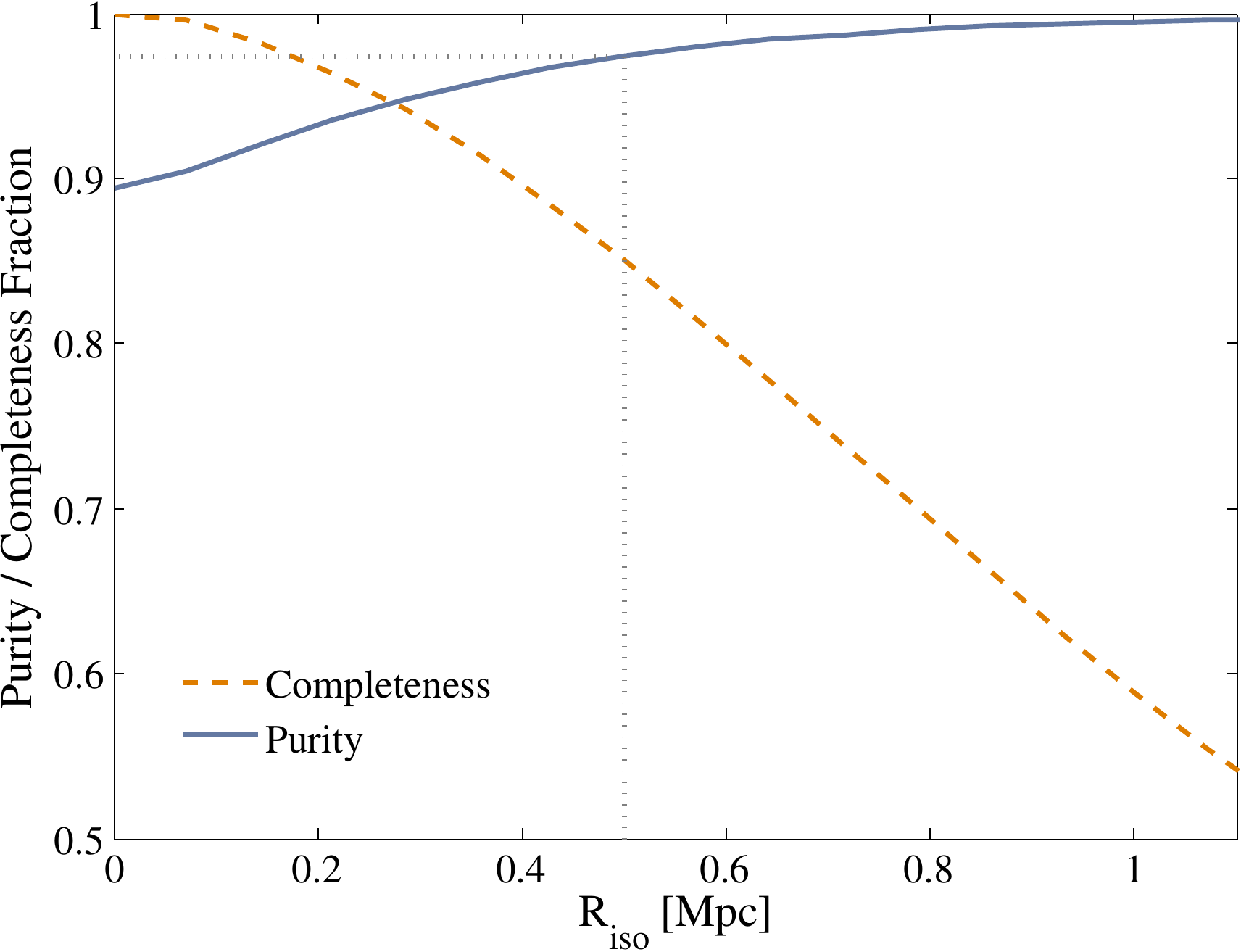}
\caption{Purity and completeness of a mock sample of MW-like galaxies
  as a function of changing isolation radius, $R_{iso}$.  The mock
  galaxies are based on halos and subhalos in a high-resolution N-body
  simulation, with similar cuts applied as to the data sample.  Over
  95\% purity is achieved for a cut of $R_{iso} = 0.5$Mpc.}
\label{fig:peter}
\end{figure}

Figure \ref{fig:peter} shows the completeness and purity of our host
sample as a function of $R_{iso}$ for Bolshoi \citep{Bolshoi}, an
N-body dark matter simulation based on the Adaptive Refinement Tree
(ART) code.  This simulation assumed flat, concordance $\Lambda$CDM
($\Omega_M = 0.27$, $\Omega_\Lambda = 0.73$, $h = 0.7$, and $\sigma_8
= 0.82$) and included $2048^3$ particles in a cubic, periodic box with
comoving side length of $357$ Mpc; the Bound Density Maxima algorithm
was used to identify halo properties \citep{Klypin99b}.  These
parameters result in halo completeness limits of about ${\rm v_{max}} >
50$ \kms, which is small enough to include the MW's massive satellites and
is well below the size of MW hosts.

For comparison with observations, we identify MW-sized halos (in the
range of $10^{12}\msol$ to $2\times 10^{12} \msol$) and compute the
projected distance $R_\ell$ to the nearest larger halo within a
redshift range of $\pm$1000 km s$^{-1}$ (projected distances are
calculated in the $x-y$ plane, and redshift-space distances are
calculated using halo positions and velocities along the $z$-axis).  A
MW-sized halo is classified as a satellite if it is within the virial
radius of a larger halo; otherwise, it is classified as a host.  Thus,
for a given $R_{iso}$, the completeness is calculated as the fraction
of MW-sized host halos with $R_\ell > R_{iso}$, and the purity is the
fraction of hosts within the set of MW-sized halos with $R_\ell >
R_{iso}$.

As one would expect, as $R_{iso}$ increases, our sample becomes more
isolated, and purity improves, but this is at the expense of rejecting
truly isolated systems from our sample.  Relatively high purity is
important to us as it impacts the relevance of our results to the MW.
However, a more complete sample will improve our resilience to
selection effects.  The choice for $R_{iso}$ thus seeks to maximize
completeness while holding impurities to an acceptably low level.  The
results of our N-body investigation are encouraging.  An isolation
radius of 500 kpc (which would count the MW as isolated, since M31 is
$\sim 700$ kpc distant) gives purity above 95\%, while still
permitting a completeness of $\sim 85\%$.  We therefore fix $R_{iso}$
at 0.5 Mpc for the core of our analysis and obtain a sample of 22,581
isolated MW analogs, extending out to $z=0.12$ (in stripe 82, with
fainter magnitude limit, we get 1946 MW analogs out to $z=0.15$).  We
show in Section \ref{sec:robust} that our results are stable upon
variation of the isolation radius, which implies that impurities at
the few-percent level do not have a significant effect.  We also
require that all potential MW analogs be at least a distance $R_{iso}$
away from the edge of the observed region.  Since we are working at
low redshift, the narrow southern SDSS stripes provide little useful
area for our analysis, and so we neglect them.

\subsection{Analogs of the Magellanic Clouds}
\label{sec:satellites}

The LMC and SMC are 2.4 and 3.8 magnitudes fainter, respectively, than
the MW in the $V$ band \citep{vandenBergh}.  Since the $V$ and $r$
bands overlap strongly, similar magnitude differences will hold in
$r$.  To find analogs of the LMC and SMC, we thus search for galaxies
around our isolated hosts within an aperture of physical size
$R_{sat}$ on the sky, and with apparent magnitudes in the range
$m_{host} + 2$ to $m_{host} + 4$. (We work in apparent magnitudes to
enable the use of the full SDSS photometric catalog, since the
magnitude difference between satellites and their hosts should be the
same in apparent or absolute values.)

The appropriate choice of $R_{sat}$ is not entirely clear.  The virial
radius of simulated dark-matter halos similar to MW is $\sim 250$ kpc
\citep{Busha10c}, so that might seem a reasonable value.  On the other
hand, the LMC and SMC are both within 100 kpc of the MW (at distances
of 50 and 63 kpc, respectively; \citealt{vandenBergh}), so if we truly
want MW analogs, we might prefer a lower value of $R_{sat}\sim 100$
kpc.  The closeness of the LMC and SMC is likely to be happenstance,
however, and an overly restrictive value for $R_{sat}$ could lead us
to underestimate the abundance of MW-MC analogs.

A further consideration in choosing $R_{sat}$ is contamination from
background objects.  Our primary analysis searches for
satellites in the photometric catalog relies on statistical
subtraction of background contamination.  As is
discussed in Section~\ref{sec:photoz}, before counting potential
satellites, we use photometric redshift information to exclude a
large fraction of the background objects.  This cut is necessarily
quite conservative, however, to avoid excluding true satellites from
our sample, so significant background remains.  The larger $R_{sat}$
becomes, the higher the background contamination becomes, and the
larger the statistical noise it induces in our results, as discussed
in Section~\ref{sec:staterrs}.  We will thus want to
keep $R_{sat}$ as small as is reasonable, to maximize our precision.
For our primary analysis, we set $R_{sat} = 150$ kpc, which strikes a
reasonable balance between these different considerations.  In
Section~\ref{sec:robust}, we explore the impact of varying $R_{sat}$
and find it to be small but not insignificant. 

Deblending is a final concern.  If a satellite is very close to its
host (either physically or in projection), the SDSS reduction pipeline
might not identify it as a separate source.  Thus, when we perform
background subtraction to obtain a spherical search volume for
satellites in Section~\ref{sec:annulusnoise} below, we are actually
considering a bead-shaped volume, with a cylinder of radius of order
$10$ kpc (roughly the radius of an MW analog) removed from the center.
Since $R_{sat}$ is much larger than the size of a typical host galaxy,
however, this cylinder represents $\la 1$\%, of the search volume (
Figure~\ref{fig:mosaic} gives a visual impression of the relative
  distances involved).  The impact of deblending on our results should
  therefore be negligibly small, and we neglect it in what follows.

\subsection{A preliminary analysis: Magellanic Clouds in the SDSS
  spectroscopic catalog}
\label{sec:spectrosat}
\begin{figure*} 
\centering
\includegraphics[width=6.5in]{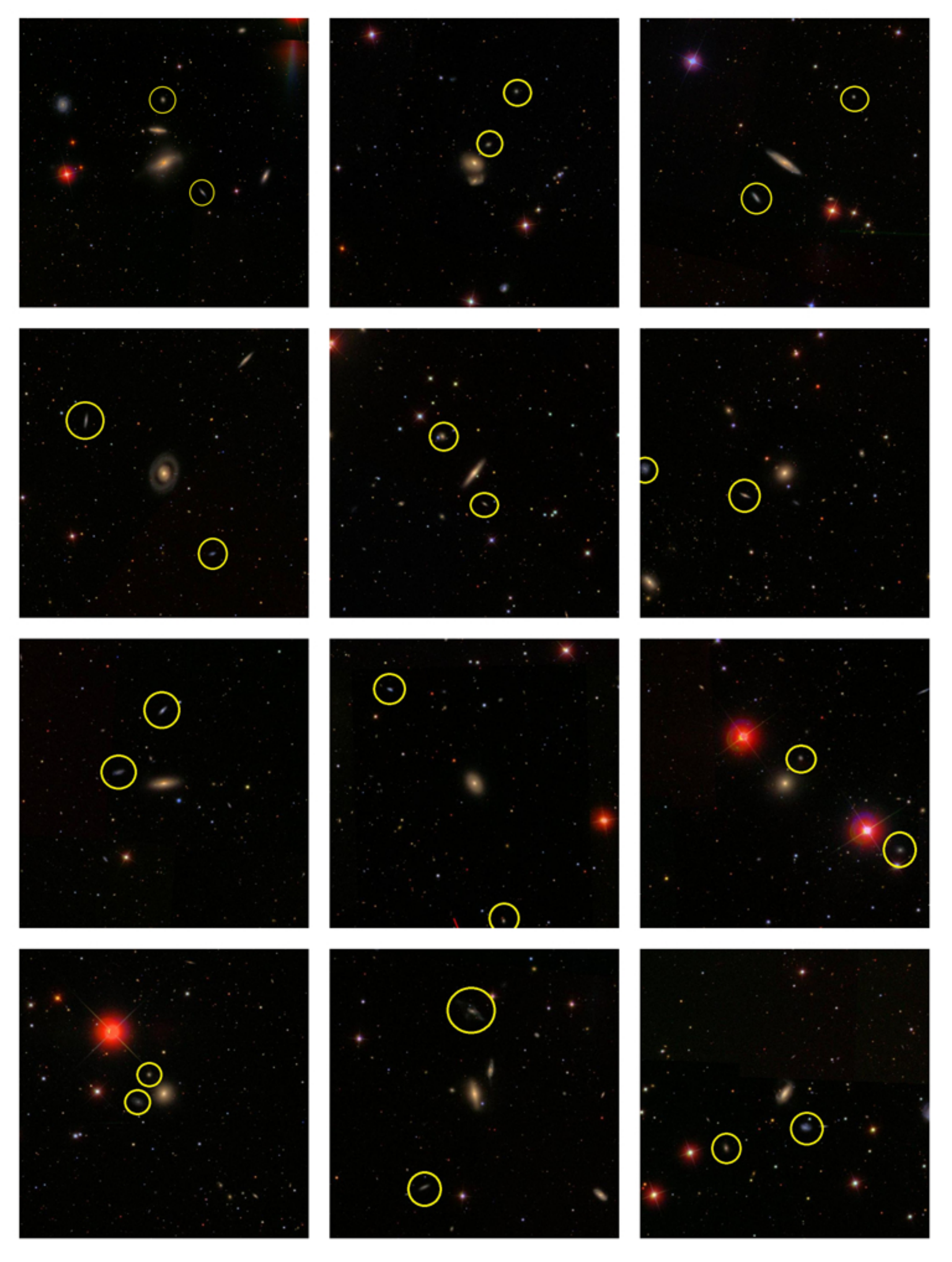}
\caption{Images of selected MW-like hosts with exactly two MC-like
  satellites in the SDSS spectroscopic catalog, identified as those
  objects within a radius of 150 kpc and within 300 \kms of the host.
  Each image is scaled to 300 physical kpc on a side, centered on the
  host galaxy.  Satellites identified as MC-like companions are
  circled in yellow.  The 1st, 2nd, 4th, and 11th images (counting
  from left to right, top to bottom) show at least one bright, close
  companion to the MW-sized host.  Image 11 shows two such objects at
  the same redshift as the central galaxy.  In each of these cases,
  the companion is recognized as a satellite of the host but is too
  luminous to meet our criteria for being an MC-like satellite.  The
  5th, 6th, 8th, 9th, and 11th images feature prominent background
  objects with spectra at dissimilar redshifts.  Background objects
  without spectra are clearly visible in every panel.  The 5th and
  12th panels exhibit fiber collisions.  The blue object next to the
  upper left MC-like satellite in panel 5, though bright enough, did
  not have its spectrum collected or analyzed, similarly, the object
  to the right of the bluer MC-like satellite in panel 12 has no
  redshift or absolute magnitude information due to fiber
  collisions. }
\label{fig:mosaic}
\end{figure*}

In this section, we generate preliminary results working exclusively
within the SDSS spectroscopic catalog.  This data set includes only the
brightest objects ($r < 17.77$) in the survey, for which spectra were
obtained.  Although these results will be subsumed by a more precise
and systematically robust result using photometrically selected
satellites, we include the brief analysis to illustrate the conceptual
simplicity of our main undertaking as well as to motivate the search
in the deeper photometric catalog.

Though the stated magnitude limit of objects in the spectroscopic
catalog is $r=17.77$, we trim the set at $r=17.60$ to avoid selection
complications near the completeness limit that arise from
recalibrations of the photometry since the main sample was selected.
This limit applies to all satellites, which implies a minimum
magnitude limit of $r=13.60$ for hosts if we allow MC-like satellites
to be four magnitudes dimmer than their hosts.

The brightest 199 members of the MW-sized galaxies selected using the
host-finding procedure outlined in Section~\ref{sec:MWhosts} (with
$R_{iso}$=0.5 Mpc, $\Delta M_{iso}$ = 0) have redshifts between 0.01
and 0.026 and $r$-band magnitudes between 12.05 and 13.60 [SDSS
units].  The search conducted around these 199 MW-like hosts
identifies as MC-like any galaxy with (1) absolute magnitude ($M_v$)
between two and four magnitudes dimmer than the magnitude of the host,
that is (2) lying within a physical projected radius of 150 kpc of
said host, and (3) has a redshift within $\Delta z_{max}$ of the
redshift of the host.  The redshift difference $\Delta z_{max}$ = 0.01
is equivalent to a $\sim 300$ \kms\ velocity dispersion, chosen as a
reasonable upper bound for the line-of-sight relative motion between a
MW host and potential satellite.

The value of $\Delta z_{max}$ also sets the uncertainty in
line-of-sight position of any potential satellite, such that the
geometry described by our limits is not a sphere centered on the
candidate host but a cylinder with the same projected dimensions. The
cylinder has a half-length of approximately 3 Mpc (as it happens, this
is roughly the correlation length of an MW-luminosity host), and any
interloper galaxies within it are indistinguishable from a true
satellite.  A systematic correction would be needed to convert this
result to expected counts within the desired spherical region with
radius $R_{sat}$ (see Section \ref{sec:systematics-lss}).  We do not
perform the correction here because the precision of our results is
already limited by our small sample size. SDSS fiber collisions will
introduce a further source of systematic error for which we would need
to correct, although this is likely to be small since the 55 arcsecond
SDSS fiber-collision radius corresponds to only $\sim 2\%$ of the search
cylinder for a typical spectroscopic host.  A more careful analysis of
the spectroscopic data in the case of LMC analogs is also in
preparation by a different set of authors (E. Tollerud et al.).  In
any case, we will derive more precise, systematically corrected
results from the photometric sample in what follows.

Here, we simply quote the result for objects with MC-like properties
found within the cylindrical redshift-space volume described above.
This method, though failing to provide the desired search geometry,
corresponds most closely to the results obtained in other
spectroscopic searches for satellites \citep[\eg,][]{zaritsky93,
  james10}, with which our results are broadly consistent.  It also
has the advantage of exact identification of individual correlated
objects (unlike our results in what follows, which are purely
statistical). Figure \ref{fig:mosaic} shows a mosaic of some likely
MW-MC-like systems identified in the spectroscopic catalog using this
procedure.  In all, from these 199 hosts, we find that 132 (or 66.3\%)
have zero, 51 (or 25.6\%) have one, 16 (or 8.0\%) have two, and none
have more than two MC-luminosity galaxies within the search cylinder.
This number-count distribution is compared with the equivalent results
from our larger photometric samples in Figure \ref{fig:stripe82}.
Even without careful systematic correction, this result stands as
qualitative confirmation of simulations \citep[\eg,][]{MBK10,
  Busha10b} showing that MW-like halos have two MC-like satellites
less than $\sim 10$ percent of the time.

\section{Methods}
\label{sec:methods}
In the last section, we motivated the need to move beyond the SDSS
spectroscopic sample to obtain statistically robust results.  The
easiest way to obtain a larger sample is to make use of the deeper
SDSS photometric catalog.  Without precise redshift information, our
analysis will depend on careful background subtraction, since
line-of-sight projection effects conflate actual satellites with
background objects.  In essence, we trade the ability to identify
individual satellites around a host for a substantial increase in
statistical power.  As discussed in Section~\ref{sec:photoz}, we can
make the task somewhat easier by using photometric redshifts to
exclude obvious background objects, but the photo-$z$s do not have
sufficient precision to identify line-of-sight interlopers on a
system-by-system basis.  We introduce here an ensemble treatment of
background subtraction performed on our expanded set of MW hosts.

Our desired result is the probability distribution function $p(S)$,
the probability that $N_{sat}$ MC-like galaxies are present within
$R_{sat}$ of an MW-sized galaxy.  We arrive at our measurement via a
four-step process, as outlined below.

\begin{enumerate}
\item We count the total number of galaxies, $T$, around each
  candidate host that meet our magnitude and projected angular
  distance ($R_{sat}$) selection criteria (see
  Section~\ref{sec:satellites}.  We use these to build a normalized
  probability distribution, the composite counts PDF,
  $p(T)$. \label{item:1}

\item We estimate the PDF of the background ``noise'' counts, $p(B)$,
  by counting galaxies meeting our satellite criteria in fields that
  do \emph{not} contain an MW-analog host.  The selection of these
  noise fields has important implications for the systematic
  uncertainty in our final result, so we take two quite different
  approaches to constructing them (see Section~\ref{sec:noise}),
  estimate the systematic errors in each case, and check that the
  results are consistent.

\item We extract the desired signal PDF $p(S)$ via deconvolution.
  Assuming signal and noise to be independent, the distribution $p(T)$
  measured in step \ref{item:1} is simply the convolution of $p(S)$
  with $p(B)$; thus, a straightforward deconvolution in Fourier space
  is all that is necessary to reconstruct $p(S)$.

\item We estimate and correct for systematic effects that arise from
  catastrophic photo-$z$ errors and from mis-estimation of the
  background contribution owing to large-scale structure, finally
  arriving at our best estimate for $p(N_{sat})$, the probability of an MW
  analog's hosting $N$ MC-like satellites.

\end{enumerate}

\begin{figure}[t!]
\centering
\includegraphics[width=3in]{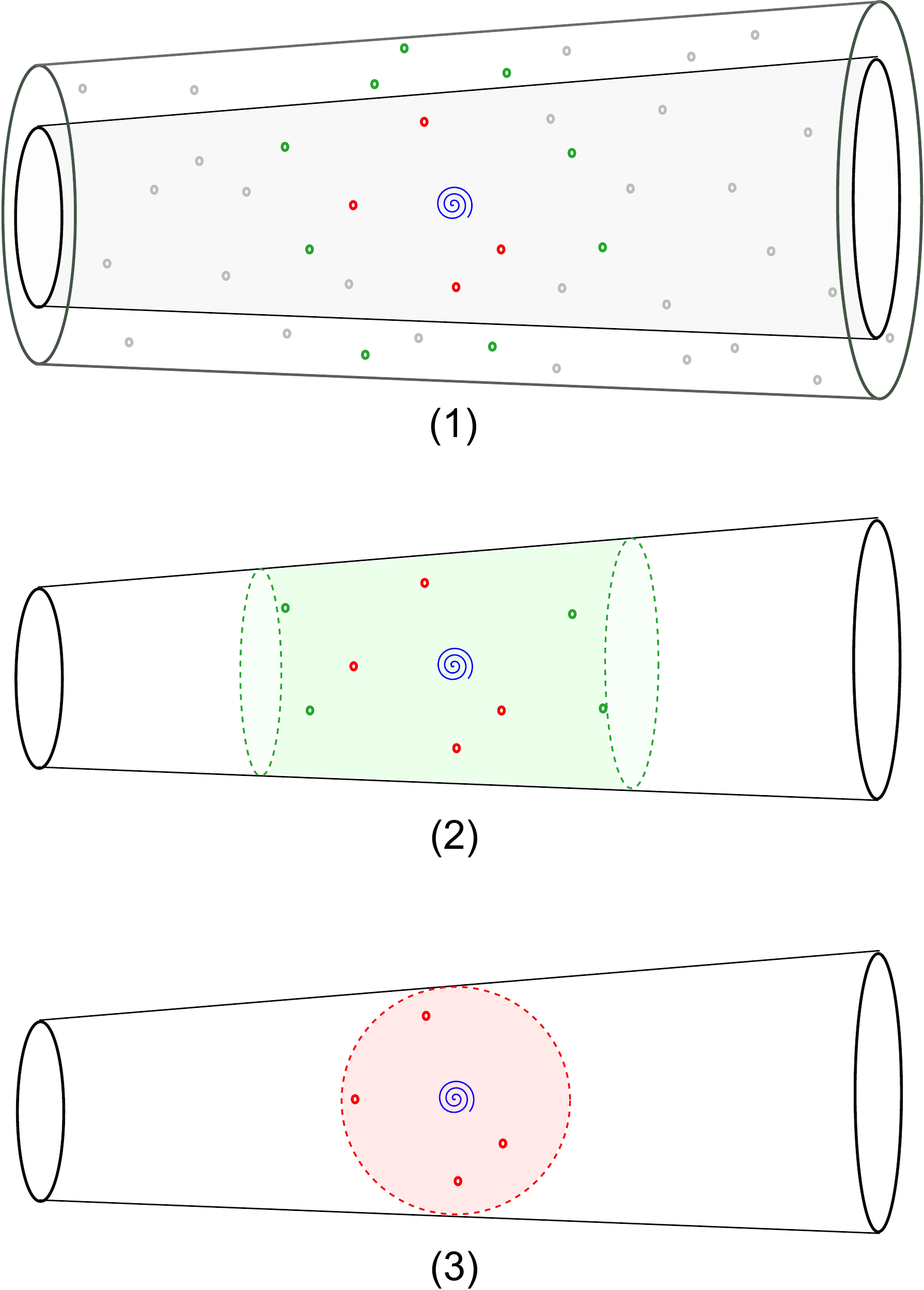}
\caption{Schematic diagrams of our background subtraction procedures.
  (1) The two volumes corresponding to the center search aperture and
  the adjacent annulus are pictured.  Red dots represent objects
  within actual physical distance $R_{sat}$ of the host, green dots
  are objects outside $R_{sat}$ but correlated with the host, and grey
  dots show random foreground and background objects. Simulations
  confirm that the amount of random and correlated background objects
  in the two volumes are approximately equal.  (2) The result for
  random background subtraction is shown.  The random background has
  been subtracted, but correlated line-of-sight structures remain,
  resulting in an effective cylindrical search volume (represented
  schematically by the green shaded region).  (3) The result of
  annular background subtraction is shown.  Both random and correlated
  line-of-sight objects have been subtracted.  This is our best
  estimate of the desired result, the number of satellites within a
  radial distance $R_{sat}$ of host.}
\label{fig:overview} 
\end{figure}

\subsection{Composite Counts}
\label{sec:totalcount}

In Section \ref{sec:MWhosts}, we presented our process for selecting
22,581 MW analogs.  Each host serves as the center of an individual
search aperture, whose angular size varies with the host redshift but
always corresponds to a transverse physical distance $R_{sat}$.  For
each aperture, a tally is made of all objects which fit our criteria
for LMC/SMC-like satellites (\eg, having apparent $r$-band fluxes
between 2 and 4 magnitudes fainter than the host, in our baseline
analysis).  The normalized histogram of these total number counts is
denoted by $p(T)$; it represents our primary measurement in this
study.

\subsection{Background estimation}
\label{sec:noise}

To estimate the PDF of background number counts, $p(B)$, we take a
similar approach to earlier studies \citep[\eg,][]{holmberg69,
  lorrimer94}.  In brief, we count galaxies that meet our satellite
selection criteria, within comparison regions on the sky that do not
contain a galaxy that meets our selection criteria for hosts (but that
otherwise meet the isolation criteria).  Previous authors taking this
approach made use of data from photographic plates, and
they wisely used comparison regions on the same plate as their host
galaxies, so the comparison regions were quite nearby the hosts.  In
our case, we have access to a large, well-calibrated photometric survey
field, so it is possible to choose comparison regions that are
arbitrarily distant from the hosts.  Since the estimation of
background noise will be the dominant source of systematic error in
this study, it is important to carefully consider the choice of
comparison fields.  We take two different approaches, which are
subject to different sources of systematic error, in order to test the
robustness of our results.

\subsubsection{Isotropic Background}
\label{sec:noisecount}

The simplest, most naive approach is to estimate the background from
random locations on the sky.  More specifically, we randomize the sky
positions of our host sample within the SDSS NGC region.  It is
important for the sake of comparison that the search is performed on
an identical distribution of aperture sizes and reference magnitudes,
however, so we do not randomize the host redshifts or luminosities.
Approximately 25,000 randomized sky positions are generated and each
is associated with a set of object properties (absolute magnitude,
apparent magnitude, redshift) belonging to a randomly chosen target
host.

These search centroids are then subjected to identical isolation
conditions as the targets with an additional constraint.  As before,
no search center may be within $R_{iso}$ of a brighter object than the
host from which the search parameters were derived.  Now, in addition,
no search center may be within $2R_{sat}$ of any MW-sized galaxy that
is within 1000 \kms of the search redshift, as we hope not to
contaminate our noise profile with signal.  The histogram of counts of
MC-like objects around these locations is then used to generate the
\emph{isotropic background} PDF, $p(B_{iso})$.

This is likely to be an underestimate of the background around our
hosts, however.  Because galaxies are clustered, regions around hosts
are generally denser than average, with a typical 
correlation length many times longer than $R_{sat}$ of this study or even
$R_{vir}$ of a galaxy.  Thus, though we have measured the random
background noise, one would expect the total noise in our
search apertures to be above random due to the contribution of
projected correlated galaxies outside our region of interest described
by $R_{sat}$. 

If no correction is made for this effect, we have in essence counted
all correlated objects within a cylinder of length roughly the
correlation length $r_0$ (see Figure~\ref{fig:overview}), which is
clearly an overestimate of the satellite population.  Fortunately, it
is straightforward to compute a correction for this systematic
undersubtraction, via integrals over the galaxy correlation function.
We derive this in Section~\ref{sec:systematics-lss} and will apply it
to our results derived using the isotropic background estimate.

It is worth noting, however, that a single correlation length
($r_0\approx 3$Mpc; \citealt{zehavi_etal:10}) along the line of sight
corresponds to $\sim 300$ \kms\ in redshift space; that is, it is the
same length as the search cylinder we used for our preliminary
spectroscopic analysis in Section~\ref{sec:spectrosat}. In other
words, the results derived from isotropic background subtraction are
roughly equivalent to our results in the spectroscopic catalog.
\emph{Even in the case of perfect spectroscopic information}, it is
necessary to account for the presence of correlated objects along
the line of sight if we wish to probe the true satellite population.
Since we did not attempt to correct our spectroscopic result in
Section~\ref{sec:spectrosat}, we will also present our uncorrected
results for isotropic background subtraction, for the purpose of
comparison.

\subsubsection{Annular Background}
\label{sec:annulusnoise}

It is also possible to estimate the random and correlated background
simultaneously and directly by placing our comparison fields very
close to the MW-analog hosts.  In particular, we can estimate the
background noise by counting galaxies that meet our satellite
selection criteria in an annulus around each host galaxy, but outside
the initial satellite search aperture, provided that the annulus has
the same projected area as the center search region.  (This technique
is similar in spirit to the one that \citealt{chen_etal:06} found to
be optimal 
for interloper removal in spectroscopic data.)  The histogram of
counts in an annulus around each host then gives the correlated
\emph{annulus background} PDF, $p(B_{ann})$.

A schematic diagram of this approach is shown in Figure
\ref{fig:overview}, panel (1).  The central column, with a sphere of
radius $R_{sat}$ cut out of it, has a slightly smaller volume than an
annulus of the same projected area, so the counts in the annulus will
tend to be slightly enhanced relative to the central cylinder.  But,
counteracting this, there is also, on average, a lower density of
correlated objects in the annulus, owing to the larger distance from
the host.  The two effects will cancel for some particular choice of
the annular radius, although it is difficult to justify \emph{a
  priori} a particular choice of this radius.

We can make some use of N-body simulations to help guide our choice.
In particular, we make use of a mock galaxy catalog generated from
abundance-matching galaxy luminosities from the low-luminosity survey of
\cite{Blanton05b} to dark matter halos in the Bolshoi simulation.  With this
catalog, we can perform identical selection cuts on MW hosts (luminosity and
isolation criteria) as we perform on observed SDSS galaxies.  Then, we may
compare the number of objects with luminosities similar to the LMC and SMC
(i.e., 2-4 magnitudes fainter than their host galaxy) in both
cylinders with the inner spheres removed, and in hollow cylinders, as
shown in panel (1) of Figure~\ref{fig:overview}.

The most natural choice for the background-estimation annulus is the
region immediately outside the search aperture, with $R_{sat}^2 <
r_{ann}^2 < 2R_{sat}^2$, which we will call Annulus I.  In this case,
tests from our mock catalogs show that the counts in the inner and
outer cylinder are roughly equal, with the counts in the annulus
possibly exceeding the counts in the search aperture, but by no more
than $\sim 10\%$.  If we move the annulus outward to $1.5R_{sat}^2 <
r_{ann}^2 < 2.5 R_{sat}^2$ (which we will call Annulus II), the
annulus counts in the simulation appear to underestimate the aperture
counts slightly, but again by no more than $\sim 10\%$.

Because the N-body models give only a rough approximation of our
measurements in SDSS, and because we would prefer not to rely too
heavily on simulations for our observational results, we do not
attempt to further optimize the radius of our search annuli.  Instead,
since our two annuli appear to tightly bracket the optimal one, we
will take the results using Annulus I to be our primary results, and
we will compare to the results using Annulus II to estimate the size
of the residual systematic uncertainty.
 
\subsection{Signal Extraction via Deconvolution}
\label{sec:deconv}

We make the assumption that the number of actual satellites to be
found around an MW-sized galaxy is unrelated to the number of
background objects which might be projected into the same aperture.
That is $S$, the signal, and $B$, the noise are independent variables.
Their sum is a third random variable, $T=S+B$.  This implies that the
probability distribution of $T$ is just the convolution of the $S$ and
$B$ PDFs:
\begin{equation}
  p(T)= p(S) \ast p(B) \equiv \sum_{S^{\prime}=0}^{T}
  p(S^{\prime})p(B^{\prime}=T-S^{\prime}), 
\label{eq:conv1}
\end{equation}  
where the $\ast$ symbol indicates convolution.  

By using the methods described in Sections \ref{sec:totalcount},
\ref{sec:noisecount}, and \ref{sec:annulusnoise}, we have precise
measurements of $p(T)$ and $p(B_{iso})$, and $p(B_{ann})$
respectively.  We are interested in $p(S_{cor})$, the probability of
encountering $S_{cor}$ LMC/SMC-like correlated galaxies within a
cylinder of radius $R_{sat}$ centered on an MW-sized host.  This is
computed by deconvolving $p(B_{iso})$ and $p(T)$.  More importantly,
we wish to obtain $p(S_{sat})$, the probability of encountering
$S_{sat}$ MC-like satellites within a sphere of physical radius
$R_{sat}$ around such a host.  This can be derived by applying a
systematic correction to $p(S_{cor})$ (see
Section~\ref{sec:systematics-lss}) or by deconvolving $p(B_{ann})$
from $p(T)$.

The deconvolutions take place in three steps. First we transform into
Fourier space using a fast Fourier transform (FFT; this is indicated
by the operator $\mathcal{F}$ below).  Then a convolution is simple
multiplication:  
\begin{equation}
\mathcal{F}(p(T)) = \mathcal{F}(p(B)) \cdot \mathcal{F}(p(S))
\label{eq:conv}
\end{equation}
By rearranging this equation we obtain
$\mathcal{F}(p(S))$, and an inverse FFT retrieves $p(S)$ in each case.
For an example (and a preview of our results), see the left-hand panel
of Figure~\ref{fig:results}.  There, the blue curve ($p(T)$) can be
obtained by 
the forward convolution of the red curve ($p(B)$) with the green curve
($p(S)$) as in Equation~\ref{eq:conv1}.  In practice, we have measured
the red and blue curves and 
deconvolved them via Equation~\ref{eq:conv} to extract the green curve.

\subsection{Use of Photometric Redshifts}
\label{sec:photoz}

As discussed below, the statistical errors in $p(S)$ depend strongly
on the typical number of background (``noise'') galaxies in the search
aperture.  We can therefore greatly improve the precision of our
results by making use of photometric redshift information to exclude
obvious background galaxies before we begin the satellite-counting
exercise outlined above.  Because photometric redshift estimates are
highly prone to catastrophically large errors---especially for faint
galaxies---we do not attempt to use photo-$z$s to identify the actual
satellites of individual hosts; instead, we merely use them to make a
conservative initial background cut.

Best-fit photometric redshift values and $p(z)$ probability
distributions are computed by \cite{cunha} for each photometric object
and are made publicly available on the SDSS DR7 webpage.  We make a
cut in the imaging catalog on best-fit photo-$z$ at some threshold
value $z_{phot,max}$ and exclude galaxies with higher photo-$z$'s from
our sample.  Because photo-$z$ estimates are prone to inaccuracies and
particularly to catastrophic errors, any cut on $z_{phot}$ will
wrongly exclude some number of galaxies that are actually satellites
at low redshift.  This will introduce a systematic undercounting of
satellites around MW-analog hosts.

\begin{figure}[tb]
\centering
\includegraphics[width=3.3in]{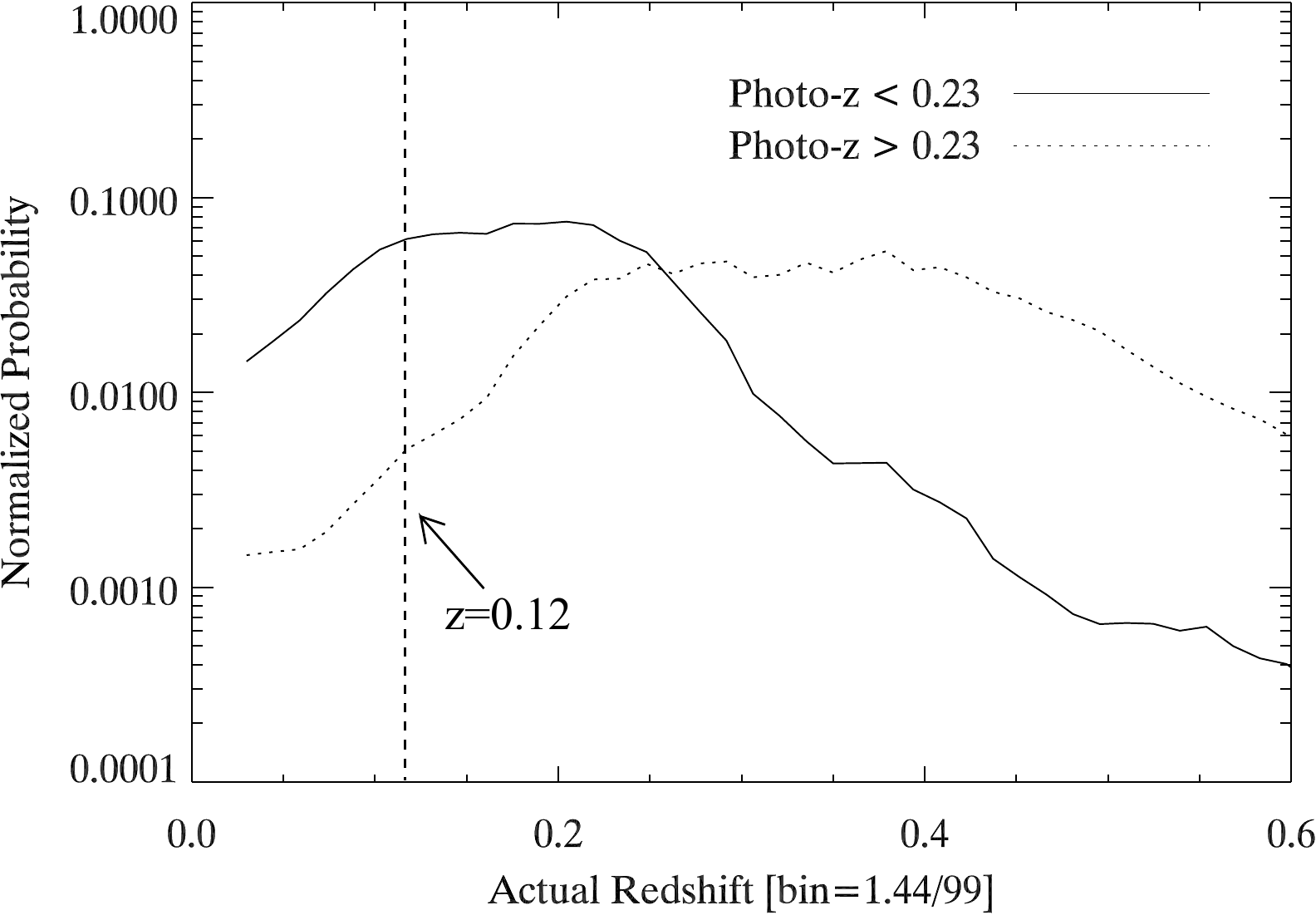}
\caption{Distribution of true redshifts for two galaxy samples divided
  by photometric redshift.  The dashed histogram indicates the average
  normalized $p(z)$ distribution from \citet{cunha} for galaxies with
  $z_{phot}$ > 0.23, and the solid histogram is the same distribution
  for galaxies with $z_{phot}$ < 0.23.  By comparing the amplitude of
  the two curves at $z<0.12$ it is possible to estimate the relative
  numbers of potential $z<0.12$ satellites that are kept (solid line)
  and excluded (dashed line) by our photo-$z$ cut.  These differ by
  roughly an order of magnitude, so the required systematic
  corrections to our satellite counts are expected to be on the order
  of $10\%$.}
\label{fig:photoz}
\end{figure}

We would like our sample of low-redshift galaxies to be as free as
possible of background objects, to reduce our statistical errors,
while also being highly complete, to minimize systematic errors.
However, increasing $z_{phot,max}$ increases the background noise and
worsens our statistical errors, while reducing $z_{phot,max}$ rejects
an increasing number of true satellites and increases our systematics.
This trade-off is shown in Figure~\ref{fig:photoz}, where we have
plotted the summed $p(z)$ distributions for a representative set of
possible MC-like galaxies with $z_{phot}$ above and below 0.23.  There
is a tail of high-photo-$z$ galaxies that are actually located at
$z<0.12$ (dotted curve) and hence ought to be considered as potential
satellites; their exclusion causes a systematic undercounting of
satellites.  There is also a large number of galaxies with
$z_{phot}<0.23$ that have true $z>0.12$ (solid line); these act as
background noise and contribute to the statistical errors.  In the
next section, we derive in detail the impact of these two sources of
error and their dependence on the photo-$z$ threshold.  We find that a
value of $z_{phot,max} = 0.23$ strikes a reasonable balance between
the purity and completeness of satellites as the resulting statistical
and systematic errors have roughly equal amplitude.

\section{Error Budget}
\label{sec:errorbudget}
\subsection{Statistical Errors}
\label{sec:staterrs}

The statistical uncertainty in our measured $p(S)$ has three sources.
The first is the overall size of our sample of MW-like host galaxies:
as our sample size increases, we expect that the precision of our
result should improve as well, owing to reduced Poisson noise.  More
specifically, the uncertainty in our measured composite counts PDF,
$p(T)$ should be purely Poisson at each value of $T$.  The second
source of error is noise from the background: if we increase our
photometric redshift cut, we increase the number of background
galaxies in our total composite counts and hence we reduce the
signal-to-noise ratio of our final measurement.  More specifically,
the isotropic background PDF, $p(B)$ is a source of noise that
propagates through our analysis in Fourier space to our final
measurements for $p(S)$.  Our sample size is large enough in our
primary analysis that we are dominated by this second source of error.

A final source of statistical error may arise from sample variance
(also sometimes called cosmic variance).  Although our observational
regions are likely numerous enough that this is not a dominant source
of error, it is possible that the variance in our composite or noise
counts exceeds simple Poisson noise.  To fully characterize the
variance in our sample, therefore, we use the jackknife technique to
estimate the errors on our measured $p(T)$ and $p(B)$.  We divide our
spatially contiguous set of MW-sized hosts into 50 subsets, each
of which may contain a different number of galaxies but occupy an
equal area on the sky.  Each iteration, a different subset is omitted,
and 49 of the 50 tiles produce a normalized PDF of counts (composite
or noise).  The result is 50 different values for each histogram bin.
Their mean is the unbiased PDF; error on the mean is approximately
$\sqrt{n-1} \cdot \sigma_i$, where $\sigma_i$ is the standard
deviation in each bin, $i$.

It is possible in principle to propagate these uncertainties
analytically through the Fourier analysis to obtain final errors on
$p(S)$. However, the scalings involved are rather non-intuitive, and
the calculation is prone to numerical instabilities when the
$p(B)$ and $p(T)$ distributions are truncated at some maximum abscissa
value, which is typically necessary.  Therefore, we propagate
uncertainties through the deconvolution using a stochastic approach.
The same FFT deconvolution is performed approximately one million
times, each time with a set of values for $p(T)$ and $p(B)$ randomly
drawn from Gaussian distributions with the means and standard
deviations found in the jackknife analysis.  To mute the effects of
ringing in the deconvolved result, we keep only those trials whose
resulting probability densities are nonnegative everywhere. The
median in each bin of the satellite counts PDF is our 
result for an MW-sized galaxy's probability of hosting $S$ =
0,1,2,3... MC-like satellites. Error bars bracket the 68\% confidence
interval.

Following this procedure, we find that the stochastic error bars derived
on $p(S)$ are much larger than the error bars estimated for $p(T)$ or
$p(B)$ in our jackknife analysis.  This indicates that the error in
$p(S)$ is dominated by background noise, rather than counting
statistics.  When we are in this regime, increasing our sample size by
a factor of order unity
will \emph{not} shrink our error bars as $\sqrt{n}$.  Instead, our
errors will thus scale roughly as the  
average signal-to-background-noise ratio, $\langle S \rangle/\langle B
\rangle$.   To improve our errors, we would need to reduce the
background, for example by making a stricter photo-$z$ cut (however,
doing this would increase our systematic errors, as discussed in the
next section).  Because we are not limited by our sample size, we have
taken an aggressive approach in our selection to excluding objects
near the edges of the NGC region.

\begin{figure}[t!hb]
\centering
\includegraphics[width=3.6in, angle=180]{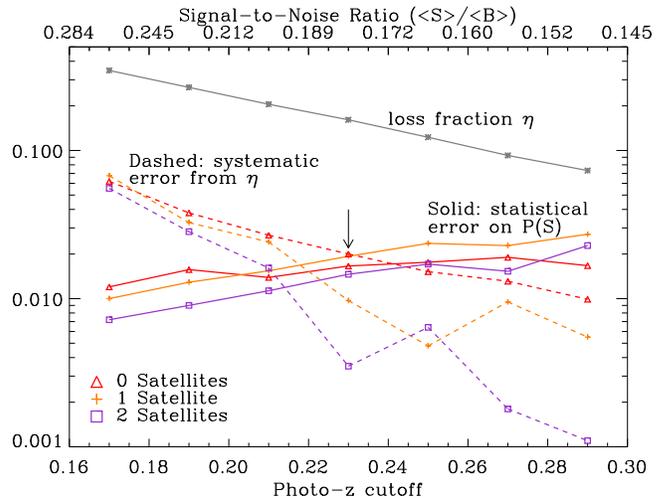}
\caption{Various sources of uncertainty in our analysis and their
  scaling with photo-$z$ cutoff.  The absolute statistical uncertainty
  on $P(S)$ increases as the photometric redshift cutoff increases and
  the average signal to noise ratio decreases.  This is shown by the
  colored solid lines, for $S=0,1,2$.  Here ``signal'' is the average
  number of MC-like satellites per Galaxy-sized host and ``noise'' is
  the background contaminating objects. A competing systematic effect
  arises from true satellites with inaccurate photo-$z$s, which are
  improperly rejected by our photo-$z$ cut.  The fraction of true
  satellites excluded in this manner is shown by the gray line, and
  the resulting systematic error is given by the colored dashed lines.
  Our adopted photo-$z$ cutoff, indicated by the arrow, is chosen to
  approximately balance these two sources of error.}
\label{fig:uncertainty}
\end{figure}  

To illustrate the scaling of our uncertainties, we compute our
errors for different values 
of $\langle S\rangle /\langle B\rangle$. We can directly obtain
$\langle T \rangle$ and $\langle B \rangle$ from our basic
number-count measurements.  Regardless of the shape of the PDFs, this
equation should then hold:
\begin{equation}
\langle S \rangle=\langle T \rangle-\langle B \rangle
\end{equation}
The most direct way of varying the signal to noise ratio is by
shifting the maximum photo-$z$ cut mentioned in Section
\ref{sec:photoz}.  Since the bulk of objects with $z_{phot} > 0.12$
are background objects, changing $z_{phot,max}$ changes $\langle B
\rangle$ while holding $\langle S \rangle$ roughly steady.  Between
$0.17 < z_{phot,max} < 0.29$, our signal-to-noise ratio varies from
approximately 0.27 to 0.15.  For our adopted value,
$z_{phot,max}=0.23$, $\langle S \rangle/\langle B \rangle = 0.18$.

Figure \ref{fig:uncertainty} plots the size of the error bars on
$p(S)$ (computed as described in Section~\ref{sec:staterrs}) against the choice of photo-$z$
cutoff and resulting $\langle S \rangle/\langle B \rangle$ for
$S=0,1,2$.  The relationship between the photo-$z$ cutoff and the
statistical uncertainty in our results demonstrates 
the need for a maximum photo-$z$ limit on the Sloan imaging catalog in
our analysis.  $\langle S \rangle/\langle B\rangle$ also varies with
the search aperture size $R_{sat}$, though this relationship is more
complicated, since the average signal (number of satellites) depends
on $R_{sat}$ as well.

\subsection{Systematic Errors}
\label{sec:systematics}

There are two primary sources of systematic error in our analysis.
First, some fraction of true satellites will be subject to
catastrophic photo-$z$ errors and thus will be wrongly rejected in our
background-exclusion cut.  This will always cause a slight
\emph{under}counting of satellites.  Second, our isotropic background
estimation in Section~\ref{sec:noisecount} assumes that
background galaxies are completely uncorrelated with
the MW-analog hosts.  Since galaxies
are in fact well known to be correlated, this technique will
lead to a slight \emph{over}counting of satellites from correlated
objects along the line of sight.  We address these
two sources of systematic error in turn below. 

\subsubsection{Photo-$z$ losses}
\label{sec:systematics-photoz}
To estimate the error caused by the photo-$z$ cut in
Sec.~\ref{sec:photoz}, we use the $p(z)$ information from \cite{cunha}
to compute a loss fraction, $\eta$.  This is the average probability
that any potential satellite object (that is, an object with the
appropriate properties and actual redshift $z<0.12$) will be cut out
of our sample owing to a catastrophic photo-$z$ error.

Photo-$z$'s of dimmer galaxies are more error-prone than those of
brighter objects since the photometric errors are larger.  Thus, an
average loss fraction must be computed 
on a sub-sample of galaxies representative of the apparent magnitude
distribution of our potential satellites.  We construct this sample
by iterating over our MW-analog hosts and, for each host, randomly selecting
1000 galaxies that are 2--4 magnitudes fainter and adding them to our
sample (this means that individual faint galaxies will appear more
than once in our sample, but this allows us to obtain the correct
magnitude distribution).  We operate on this set by dividing it
into two, those objects with best-fit photo-$z$ above the
threshold $z_{phot,max}$, those objects with best-fit photo-$z$ below
this cut
(see Figure~\ref{fig:photoz}).

Then, if $L$ is the set of all potential satellites with best-fit photo-$z$
below $z_{phot, max}$, the loss fraction is
\begin{equation}
\label{eq:cond}
\eta=p(g \not\in L | z_g < 0.12)
\end{equation}
Since we do not have direct access to the actual redshifts, $z_g$, of
individual photometric objects, we use Bayes's Theorem to rewrite the
conditional probability in a more accessible form:
\begin{equation}
\eta=p(g \not\in L | z_g < 0.12) = \frac{p(z_g < 0.12 | g \not\in L)
  p(g \not\in L)}{p(z_g < 0.12)}. 
\end{equation}

In Figure \ref{fig:photoz}, $L$ is represented by the solid line, and
the set of all other objects, which we can call $M$, is represented by
the dotted line, so $p(z_g < 0.12 | g \not\in L)$ is the integral
under the dotted line between $z=0$ and $z=0.12$.  $p(g \not\in L)$ is
the ratio between the size of $M$ and the size of the full set $M
\cup L$, and $p(z_g < 0.12)$ is the integral from $z=0$ to $z=0.12$
of the normalized $p(z)$ distribution of $M \cup L$.

\begin{figure}[t!]
\centering
\includegraphics[width=3.3in]{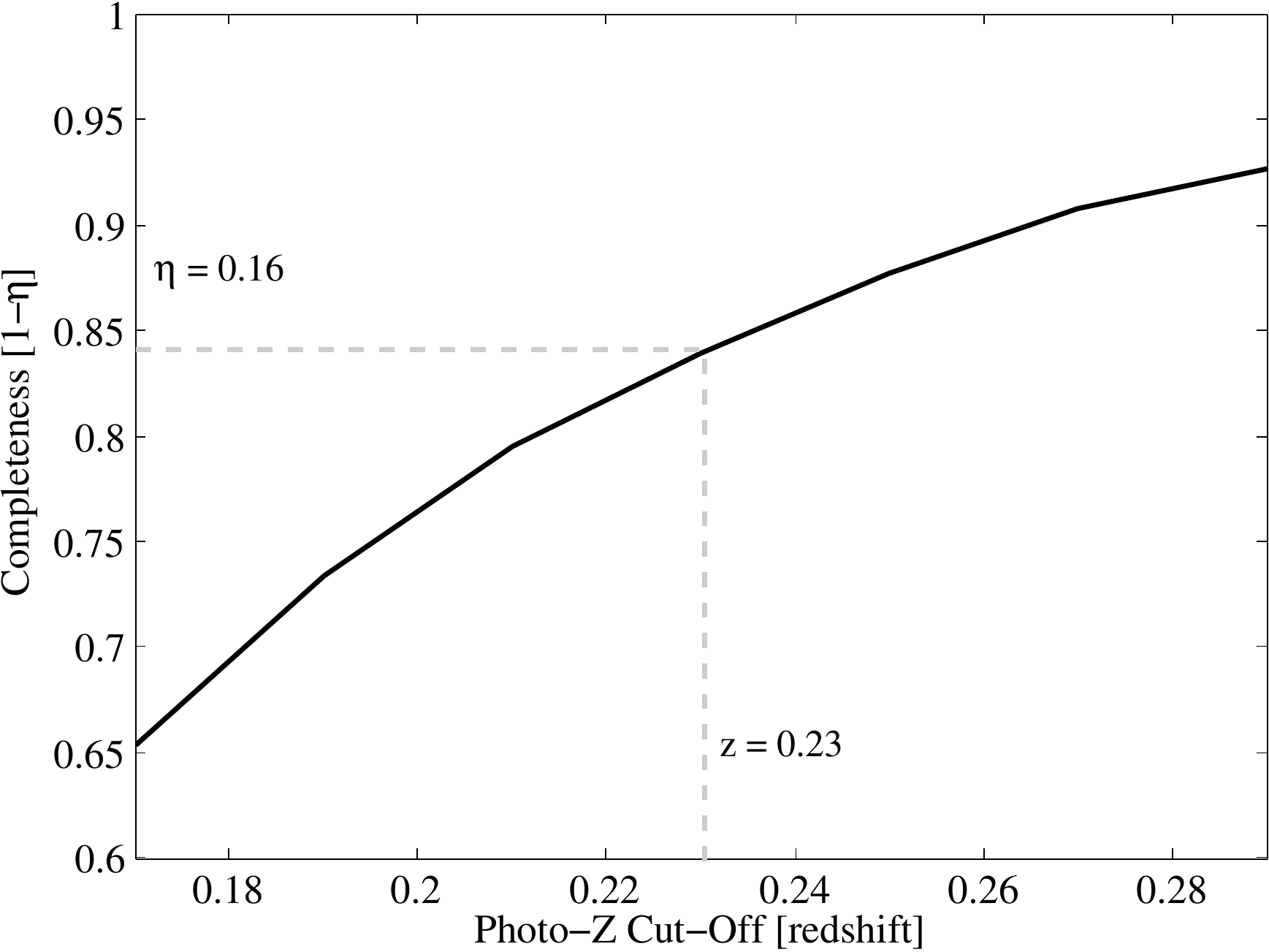}
\caption{Completeness of $z<0.12$ objects (\ie the fraction of
  $z<0.12$ objects with $z_{phot} < z_{phot,max}$) as the maximum photo-$z$ cut-off,
  $z_{phot,max}$, is increased.  }  
\label{fig:photoz_var}
\end{figure}

The galaxy completeness after the photo-$z$ cut, $1-\eta$, is plotted
against $z_{phot,max}$ in Figure \ref{fig:photoz_var}.  For our
primary results searching in the main Sloan imaging catalog with
$z_{phot,max}$=0.23, $\eta$=0.16.  We thus expect the impact of this
first source of systematic error to be at the $\sim 10$ percent level.

Given an estimate for $\eta$, we can compute a straightforward
correction for the systematic error from photo-$z$ losses. We relate
$p_{meas}(S)$, the measured probability distribution for MC-like
satellites around MW-like hosts, to $p_{true}(N)$, the actual
distribution, applying the overall loss-fraction, $\eta$, uniformly as
a loss probability for each satellite and including appropriate
combinatorial factors.  For a galaxy with $N$ actual satellites, the
probability that exactly $m$ satellites will be lost is,
\begin{equation}
p_{loss}(m|N)=\eta^{m}(1-\eta)^{N-m}\binom{N}{m}.
\end{equation}
Then the measured satellite PDF is related to the true PDF by
\begin{equation}
p_{meas}(S) = \sum_{m=0}^{\infty}p_{true}(N \equiv S+m)p_{loss}(m|N).
\label{eqn:losscorrection}
\end{equation}
This equation corresponds to a formally infinite system of equations,
one for each value of $N$. Since $p_{meas}(S)$ and (presumably)
$p_{true}(N)$ approach zero as $N$ increases, however, we may solve
for $p_{true}(N)$ by truncating at some appropriately large values of
$N$ and $S$ (chosen to be 15, well beyond where the average value is
zero). This gives a tractable system of equations, which we then solve
to obtain a result corrected for photo-$z$ losses.

\subsubsection{Large-scale structure effects}
\label{sec:systematics-lss}

To estimate the impact of correlated structure along the line of
sight, we would like to compute an analogous quantity to the loss
fraction, $\eta$---we will call it the \emph{boost fraction},
$\zeta$---that quantifies the fraction of our satellite counts that
can be attributed to line-of-sight structure after we have made an
isotropic background correction.  To do this, we make use of the
galaxy autocorrelation function $\xi(r)$, which quantifies the excess
probability above random of finding a galaxy some distance $r$ from
another and which is well measured in the local universe.  Strictly
speaking, since we are considering the correlations between two
different galaxy populations, we should use the cross-correlation
function of these two samples, but given that both MW-sized objects
and LMC-sized objects should be roughly unbiased tracers of the
dark-matter distribution \citep[\eg,][]{zehavi_etal:10}, their
cross-correlation and autocorrelation functions will be approximately
equal.

In particular, we can make use of the projected correlation function
$w_p(r_p)$, which is given by integrating $\xi(r)$ along the line of
sight.  This function gives the excess probability (above random) of
finding a galaxy at a projected distance $r_p$ away from another on
the sky.  At $r_p<R_{sat}$, the dominant contribution to $w_p(r_p)$ is
from true satellite galaxies, but there will also be some contribution
from unbound galaxies along the line of sight.  We can estimate the
size of this contribution by integrating $\xi(r)$ along the line of
sight, \emph{excluding a sphere of radius} $R_{sat}$ \emph{around the
  origin} and comparing this to the full $w_p(r_p)$.  Following
\citet{DavisPeebles83}, the modified projection we want is
\begin{equation}
\widehat{w_p}(r_p) = \int_{r_{min}(r_p)}^{\infty} r\; dr\; \xi(r)\; (r^2 -
r_p^2)^{-1/2},
\end{equation}
where the lower limit of integration is 
\begin{equation}
r_{min}(r_p) = \left(r_p^2 + \sqrt{R_{sat}^2 - r_p^2}\right)^{1/2}
\end{equation} 
and defines the sphere
within which we wish to count satellites.  This can be integrated
numerically for a given choice of $\xi(r)$.  

If we let $R_{sat} \rightarrow
0$ and assume a power-law form for the correlation function, $\xi(r) =
(r/r_0)^{-\gamma}$, we obtain the well-known analytic formula for $w_p(r_p)$, 
\begin{equation}
w_p(r_p) = r_p\left(\frac{r_p}{r_0}\right)^{-\gamma}
\Gamma\left(\frac{1}{2}\right)
\Gamma\left(\frac{\gamma-1}{2}\right) \left/
\Gamma\left(\frac{\gamma}{2}\right) \right. .
\end{equation}
Then, by integrating both $\widehat{w_p}$ and $w_p$ out to $R_{sat}$,
we can compute the probability that a satellite candidate is
correlated with its putative host, beyond what is accounted for by our
isotropic background correction, but is not actually within $R_{sat}$.
This probability is the boost fraction $\zeta$:
\begin{equation}
\zeta = \int_0^{R_{sat}} \widehat{w_p}(r_p)\, dr_p \left/
\int_0^{R_{sat}} w_p(r_p)\, dr_p. \right.
\end{equation}
Assuming $\gamma = 1.8$ (approximately the value for galaxies dimmer
than L*; \eg, \citealt{zehavi_etal:10}), when $R_{sat} = 150$ kpc we
obtain $\zeta = 0.21$.

The probability that exactly $n$ correlated galaxies will be counted
along the line of sight is then
\begin{equation}
p_{boost}(n) = \zeta^n (1-\zeta).
\end{equation}
The second factor ensures that the probability
distribution is normalized (since the sum over $n$ is a geometric
sequence); it accounts for the probability of having zero correlated
line-of-sight systems.  The systematic correction for correlated
structure can then be derived as in the previous section, by relating
the measured PDF to the true PDF:
\begin{equation}
p_{meas}(S)=\sum^{S}_{n=0} p_{true}(N=S-n) p_{boost}(n).
\label{eqn:boostcorrection}
\end{equation}
We can solve this as before by truncating the formally infinite system
of equations at suitably large $S$ such that $p_{meas}(S)$ vanishes.
In practice, we first compute the correction for photo-$z$ losses from
Equation~\ref{eqn:losscorrection} and then we compute the boost
correction using the results of that calculation.  This ensures that
we account correctly for correlated non-satellite galaxies that were
lost to photo-$z$ failures.

Before moving on, we make note of a possible inaccuracy in the
analysis in this section.  We have assumed that $\zeta$ does not
depend on the true number of satellites, $N$. However, since $\zeta$
depends on the bias of the hosts, this may not be completely correct.
One might imagine that the satellite population depends, to some
extent, on the formation epoch of the hosts (since hosts forming
earlier have more time to disrupt or merge with their satellites).
Galaxy biasing is also known to depend on formation epoch (the
so-called ``assembly bias''), and this effect is at the $\sim 20\%$
level for halos like the MW \citep{Wechsler06}.  In fact,
\citep{Busha10b} show explicitly that there is some dependence of the
satellite number on environment in this mass regime.  $\zeta$ will
depend linearly on the host bias via the host-satellite cross
correlation function.  However, including this effect would complicate
our analysis substantially: we would no longer be able to separate
Equations~\ref{eqn:losscorrection} and \ref{eqn:boostcorrection}, and
we would have to write them as a double sum, yielding a much more
complicated system of equations.  Because the effect is of order
$20\%$ on top of a boost fraction that is of similar order, we treat
it as a second-order correction and neglect it.  

\section{Results}
\label{sec:results}

\subsection{Primary Results}
\label{sec:primaryresults}
To compute our main results we use the parameters $R_{iso}$=0.5 Mpc,
$\Delta M_{iso} = 0$ (\ie, only rejecting galaxies as non-isolated if
they have a \emph{brighter} companion), $\Delta M_{sat} = 2$
(searching satellites 2-4 magnitudes dimmer than host), $R_{sat}$=150
kpc, and $z_{phot,max}$=0.23.  The maximum photo-$z$ value is chosen
to yield random errors that are greater than or similar to the
systematic errors from photo-$z$ losses (see Figure
\ref{fig:uncertainty}), as discussed in
Section~\ref{sec:systematics}. We note that our isolation and
satellite-search parameters would select the MW-LMC-SMC system, since
our nearest bright neighbor, M31, is 0.7 Mpc distant, and the LMC and
SMC are both well within 150 kpc of the MW.  In what follows, we will
vary these parameters to check the robustness of our results; we find
the satellite counts to be relatively insensitive to the choice of
parameters.

\begin{figure*}[tb]
\centering
\includegraphics[width=3.5in]{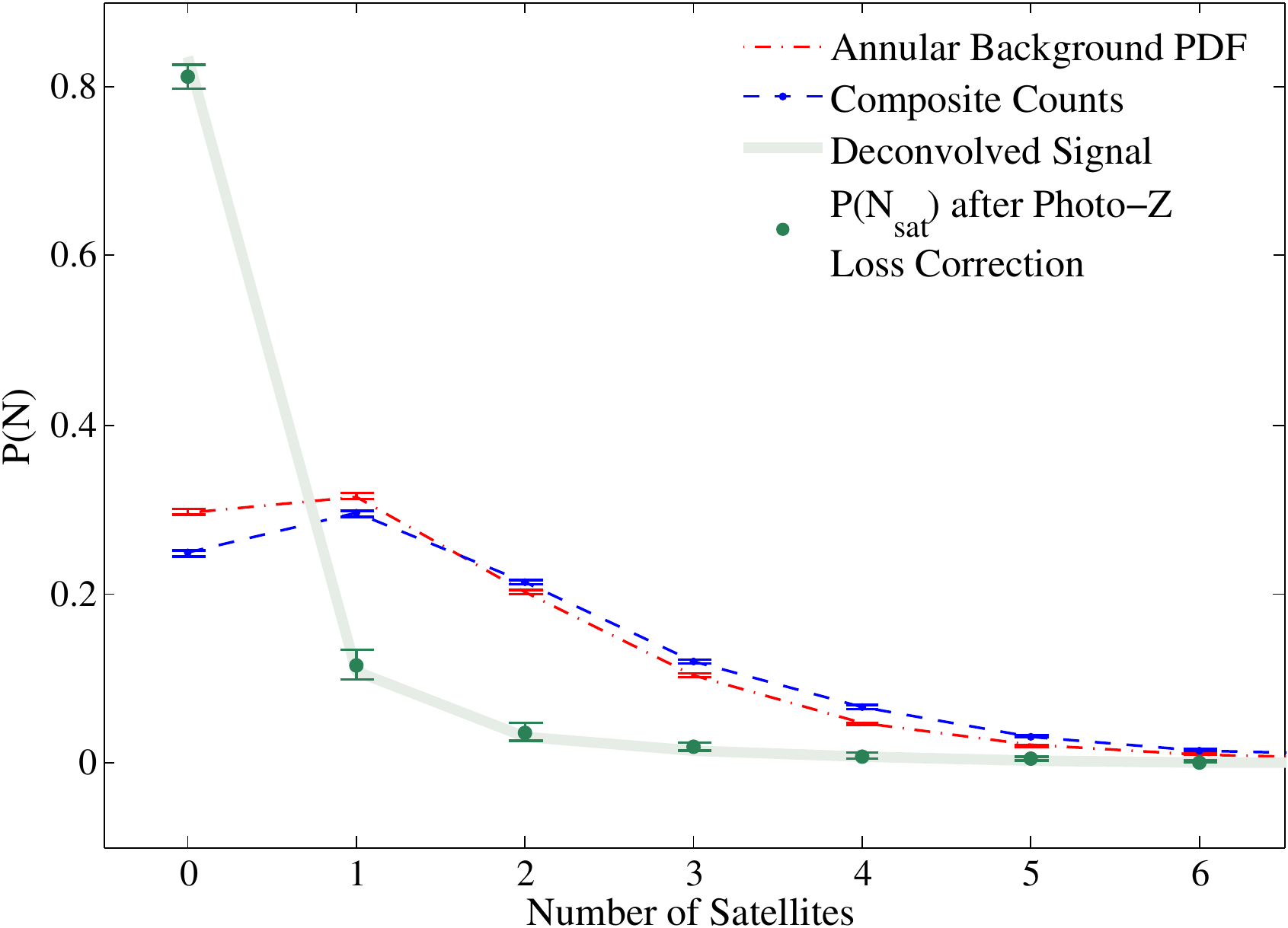}
\includegraphics[width=3.5in]{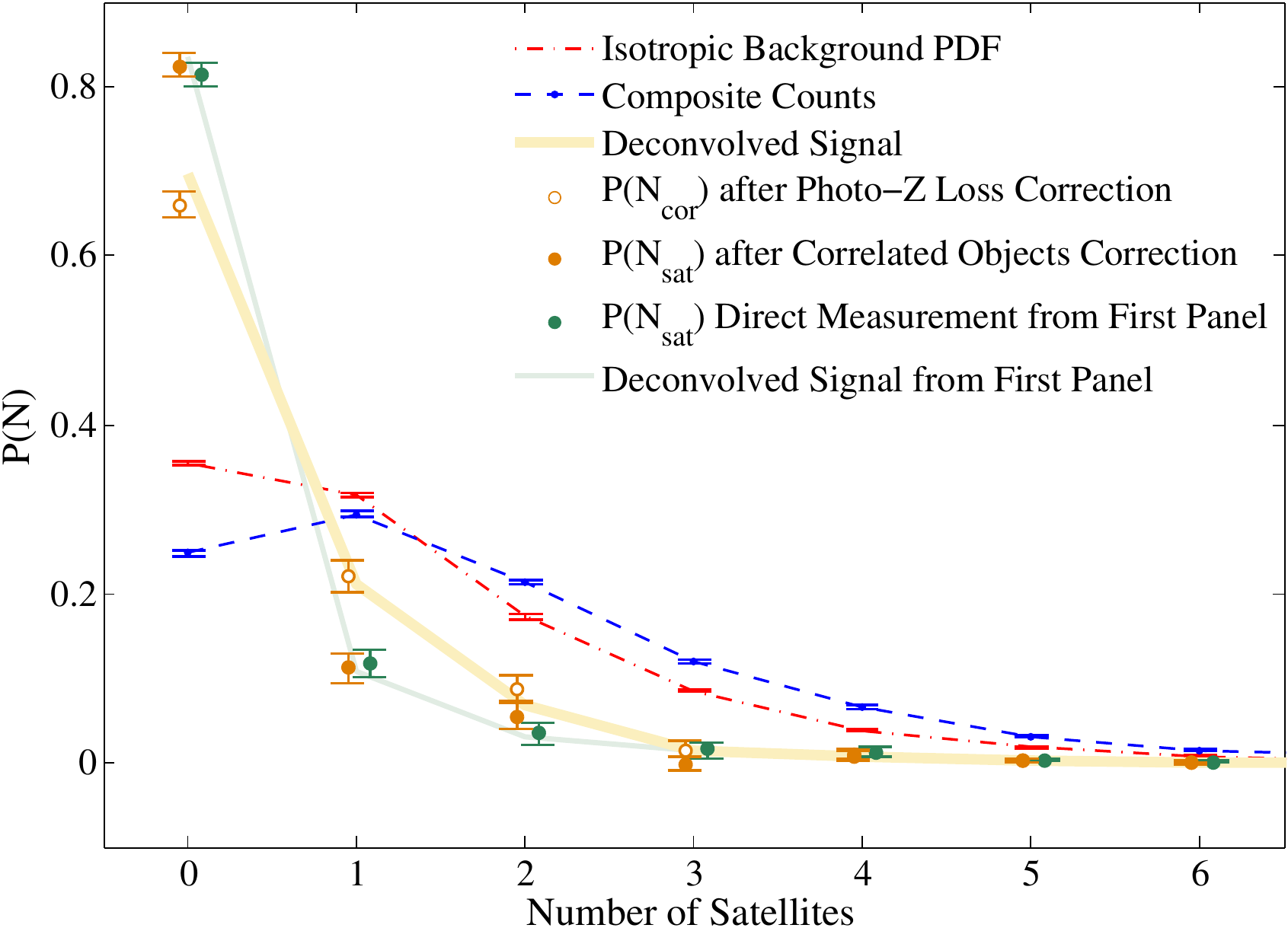}
\caption{Probability that an MW-sized galaxy hosts $N_{sat}$
  LMC/SMC-like satellites or $N_{cor}$ line-of-sight correlated
  structures.  \emph{Left:} Our primary result, $p(N_{sat})$ (solid
  green points) computed using annular background estimation, along
  with the various steps involved in deriving this result, described
  in Section~\ref{sec:methods}.  The blue curve shows the composite
  counts PDF, $p(T)$ and the red curve shows the background counts PDF
  estimated from annuli around each MW analog.  The green curve is the
  deconvolution of these two PDFs, $p(S)$, and the green data points
  are the values for $p(N_{sat})$ after correction for photo-$z$
  losses.  \emph{Right:} Similar curves are shown, but now for the
  case of isotropic background estimation.  Now the orange curve is
  the deconvolved $p(S)$ and the open orange points are $p(N_{cor})$
  after correction for photo-$z$ losses.  Solid orange points show our
  best estimate for $p(N_{sat})$ in this case, after computing a
  correction for correlated structure along the line of sight.  We
  find that these results compare favorably to the results from left
  panel (green points and curve), which suggests that our results are
  robust.  The host and satellite selection parameters used in this
  analysis are $R_{iso}$=0.5 Mpc, $R_{sat}$=150kpc,
  $z_{phot,max}$=0.23 ($\eta$=0.16), and $\Delta M_{iso}$=0. }
\label{fig:results}
\end{figure*}

In Figure \ref{fig:results} and Tables \ref{tab:results-sphere} and
\ref{tab:results-cyl} we report the percentage of MW-sized galaxies
with $N$ satellites or correlated objects centered on the host.  $N$
takes on integer values, and is labeled $N_{cor}$ for the result
accomplished through isotropic background subtraction and $N_{sat}$
for the result achieved through annular background subtraction.

Our annular background-subtraction technique gives our best estimate
for the counts of dwarf satellites within a sphere of radius $R_{sat}$
centered on each MW-sized host.  We find the probability of there
occurring $N_{sat}$ = 0, 1, and 2 bound MC-like satellites to be (81.4
$\pm$ 1.5) \% , (11.6 $\pm$ 1.8) \%, and (3.5 $\pm$ 1.4) \%
respectively, after adjustment for systematic errors arising from
catastrophic photo-$z$ failures (see
Section~\ref{sec:systematics-photoz}).  The measured values and
systematic corrections are tabulated in Table~\ref{tab:results-sphere}
and plotted in the left-hand panel of Figure~\ref{fig:results} (green
curve and data points).  Also plotted in that figure are the composite
counts PDF $p(T)$ and the background PDF $p(B)$ (blue and red curves,
respectively) which are the curves that have been deconvolved to yield
the measured satellite counts.

\begin{deluxetable}{lccc}
\tablewidth{0pt}
\tablecaption{
    Percentage of MW-luminosity host galaxies with N LMC/SMC luminosity satellites within a sphere of radius 150kpc, 
    for N=0-6}
\tablehead{
\colhead{Satellite} & \colhead{Measured $\%$} & \colhead{Systematic Loss} & \colhead{Annulus Systematic} \\
\colhead{Counts}  & \colhead{of MW analogs} & \colhead{Adjustment} & \colhead{Uncertainty\tablenotemark{a}}
}
\startdata
Zero &	$83.4^{+1.5}_{-1.4}$  &  -2.0  & -4.2 \\
One	&    $ 10.8^{+1.8}_{-1.6}$  &  +0.8 & +2.6 \\
Two	&    $ 3.1^{+1.3}_{-1.5}$   &  +0.4 & +1.6 \\
Three	&   $ 1.4^{+0.9}_{-1.0}$  &  +0.2 & +0.1 \\
Four	&   $0.7^{+0.6}_{-0.5}$  &  +0.4 & +0.2 \\
Five	&   $0.1^{+0.2}_{-0.1}$  &  +0.2 &  +0.1 \\
Six	&   $0.1^{+0.2}_{-0.1}$   & -0.1 & +0.1 
\enddata
\tablenotetext{a}{This is our estimate for the \emph{maximum} additional correction that
might be required to account for having chosen a non-optimal annulus
for background estimation.}
\label{tab:results-sphere}
\end{deluxetable}

\begin{deluxetable}{lccc}
\tablewidth{0pt}
\tablecaption{
Percentage of MW-luminosity host galaxies with N LMC/SMC luminosity
correlated objects within a cylinder of radius 150kpc,  for N=0-6}
\tablehead{
\colhead{Correlated} & \colhead{Measured $\%$} & \colhead{Systematic Loss} & \colhead{Systematic Boost}\\
\colhead{Objects} & \colhead{of MW analogs} & \colhead{Adjustment} & \colhead{Adjustment}
}
\startdata
Zero &	$69.7^{+1.6}_{-1.3}$  &  -3.8 & +16.5\\
One	&    $ 21.1^{+1.7}_{-1.9}$  &  +1.0 & -11.1\\
Two	&   $ 6.8^{+1.5}_{-1.5}$   &  +2.1 & -3.4\\
Three	&   $ 1.3^{+1.0}_{-0.7}$  &  +0.2 & -1.8\\
Four	&   $0.6^{+0.7}_{-0.5}$  &  +0.3 & -0.1\\
Five	&   $0.1^{+0.3}_{-0.1}$  &  +0.1 & -0.1\\
Six	&   $0.1^{+0.2}_{-0.1}$   & -0.0 & -0.1
\enddata
\label{tab:results-cyl}
\end{deluxetable}

We derived these results using the comparison region immediately
outside our search aperture that we called Annulus I in
Section~\ref{sec:annulusnoise}. As we discussed in that Section, this
may yield a very slight overestimation of the background, according to
our tests in simulations.  To quantify the potential size of this
residual systematic error, we also compute our results using Annulus
II (which simulations suggest is likely to yield a very slight
\emph{under}estimate of the background).  We take the difference
between these two results to be an estimate for the \emph{maximum}
remaining systematic error in our primary results; we report this in
the final column of Table~\ref{tab:results-sphere}.

The isotropic background correction yields counts of MC-like dwarf
galaxies correlated with the host within a cylinder around the host 
with radius $R_{sat}$ and effective half-length of roughly the
correlation length of unbiased mass tracers.  For $N_{cor} = 0, 1$ and
$2$, we find probabilities 
(64.6 $\pm$ 1.5) \%, (22.8 $\pm$ 1.8), and (9.7 $\pm$ 1.5)
\%, respectively.  These numbers
have also had the systematic correction for photo-$z$ loss applied; the
measured numbers and the corrections are tabulated in
Table~\ref{tab:results-cyl} and plotted in the right-hand panel of
Figure~\ref{fig:results} (thick orange curve and open data points),
along with the composite and background 
PDFs.  As discussed in Section~\ref{sec:noisecount}, this result is
the most directly comparable to satellite counts measured in redshift
space, if interlopers have not been accounted for, so we present it
here for comparison to our results in Section~\ref{sec:spectrosat}.
   
In Section \ref{sec:systematics-lss}, we developed a further
systematic correction to allow us to remove the effects of correlated
line-of-sight structures from this result. We compute this correction
for the results of the isotropic background subtraction, and we give
the results in the final column of Table~\ref{tab:results-cyl}.  We
also plot the corrected probabilities in the right panel of
Figure~\ref{fig:results} (solid orange points) and compare to the
results of our annular background correction (green points), The good
agreement between these two approaches gives us confidence that our
methods are robust.  We note that a similar systematic boost
correction could also be usefully applied to any future spectroscopic
satellite searches to account for correlated interlopers.

Our results compare favorably with data from recent
high-resolution numerical N-body simulations, such as the Millennium-2
simulation \citep{MBK10} and the Bolshoi simulation.  The latter
agreement will be discussed in more detail in a companion paper to
this one \citep{Busha10b}.  
It is also worth mentioning that none of our measurements of
$p(N_{sat})$ is consistent with a Poisson distribution with an
expectation value of $<N>=0.3$.  A detailed discussion of this
can be found in the companion paper \citep{Busha10b}.

\subsection{Satellite populations as a function of host-galaxy color}

These results suggest that the MW, with two
large, close satellites, is not a typical galaxy for its luminosity.
Since the MW is a blue, star-forming galaxy, we can take the analysis
one step further and investigate whether the 
number of satellites is a function of galaxy color. This may be quite
worthwhile, since the SDSS sample is dominated by red galaxies, and
this could complicate the implications of our study for the MW.
Galaxy colors in  the local universe are well known to be  
bimodal \citep{Strateva01}, and we can cleanly divide our sample into red
and blue objects by cutting at $u-r = 2.4$.

We repeat our analysis for the red and blue samples separately, using
annular background estimation, and we find no statistically
significant difference between the satellite statistics of the two
sets.  The results are provided in Table \ref{tab:redblue}, where
systematic adjustments for photo-$z$ losses have been applied to the
numbers given (the adjustments were not applied in
Tables~\ref{tab:results-sphere} and \ref{tab:results-cyl}). This
result appears to be at odds with work by \citet{lorrimer94} and
\citet{chen:08},  who
found more satellites around early-type galaxies, on average, than
around late-types.  However, those studies considered a wider range
in host luminosity than we have done here, and so it is likely that the
early-type samples were 
skewed toward brighter magnitudes than the late-type sample.
The fact that we find no significant difference in our larger sample,
which is limited to a narrow range in host luminosity, suggests that
the earlier results may have mainly uncovered a trend with host-galaxy
luminosity, rather than galaxy type.

It is reasonable to wonder how our results change if we divide the
\emph{satellite} population by color, especially since the MCs are both blue,
star-forming galaxies.  However, since we do not have very accurate
photo-z estimates for faint SDSS galaxies, we also lack good
$k$-corrections for these objects, and so their absolute colors are
uncertain.   In order to produce robust and reliable results on the
color dependence of the satellite population, more
accurate photo-z estimates would be required.  We therefore do not
attempt to perform this test here.


\begin{deluxetable}{lccc}
\tablewidth{0pt}
\tablecaption{Satellite statistics of red- and
      blue-sequence MW-sized galaxies, using annular background
      estimation and  after systematic adjustment.}
\tablehead{
\colhead{Number of} & \colhead{Red Galaxies} & \colhead{Blue Galaxies} & \colhead{Average} \\
\colhead{Satellites} & \colhead{ $P(N_{sat})$} & \colhead{$P(N_{sat})$} & \colhead{$P(N_{sat})$} 
}
\startdata
Zero &   82.0 & 81.2 & 81.5  \\
One	&    11.6 & 12.5 & 11.7  \\
Two	&    2.6 & 3.5 & 3.5  \\
Three &	2.1 &  0.5&  1.5  \\
Four	&   0.8 & 1.3 & 1.1  \\
Five	&   0.3 & 0.3& 0.3  \\
Six	&   0.0 & 0.0& 0.0 
\enddata
\label{tab:redblue}
\end{deluxetable}

\subsection{Robustness of the Results}
\label{sec:robust}

\subsubsection{Varying the selection and search criteria}
\begin{figure*}[t!]
\centering
\includegraphics[width=3.3in]{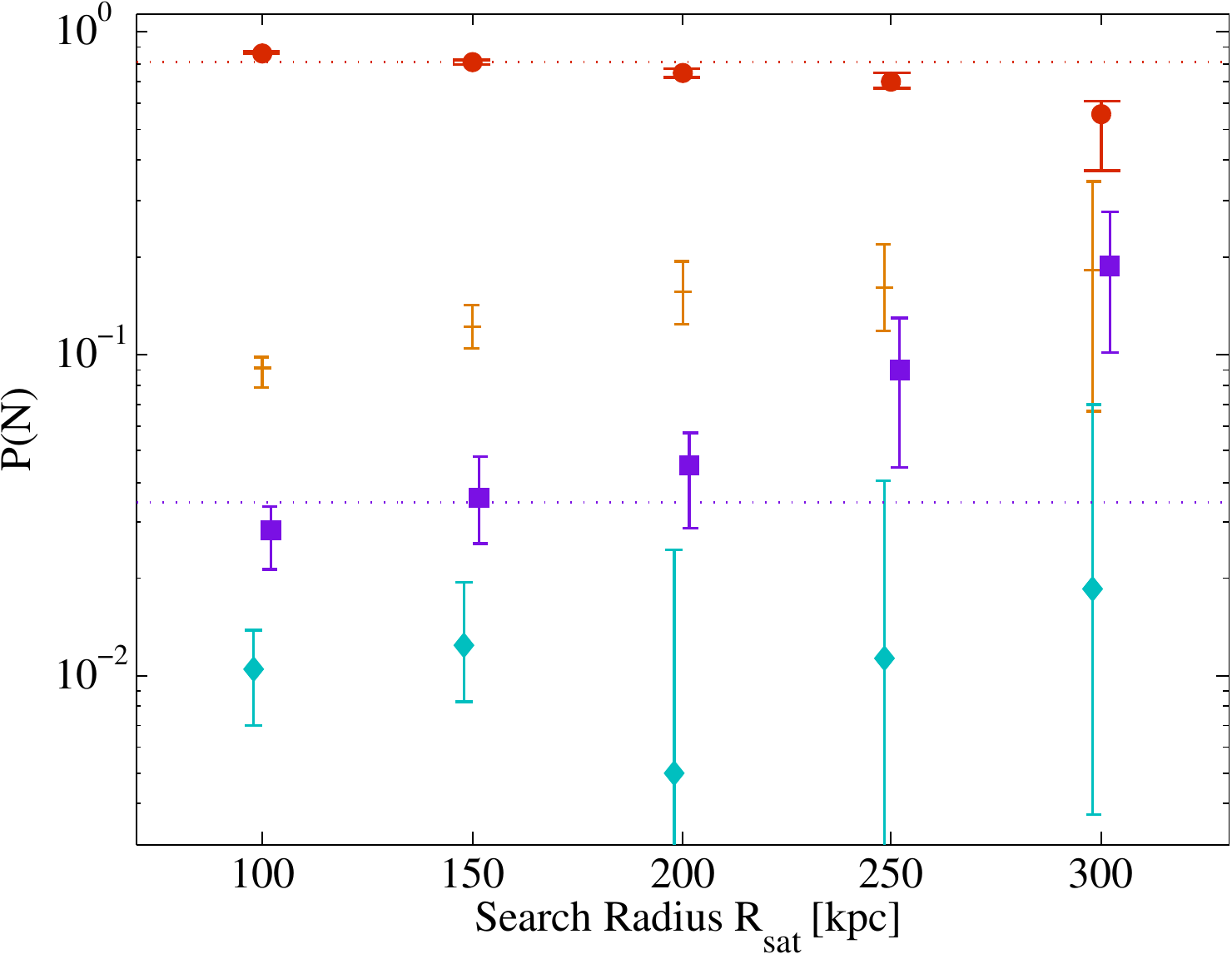}
\includegraphics[width=3.3in]{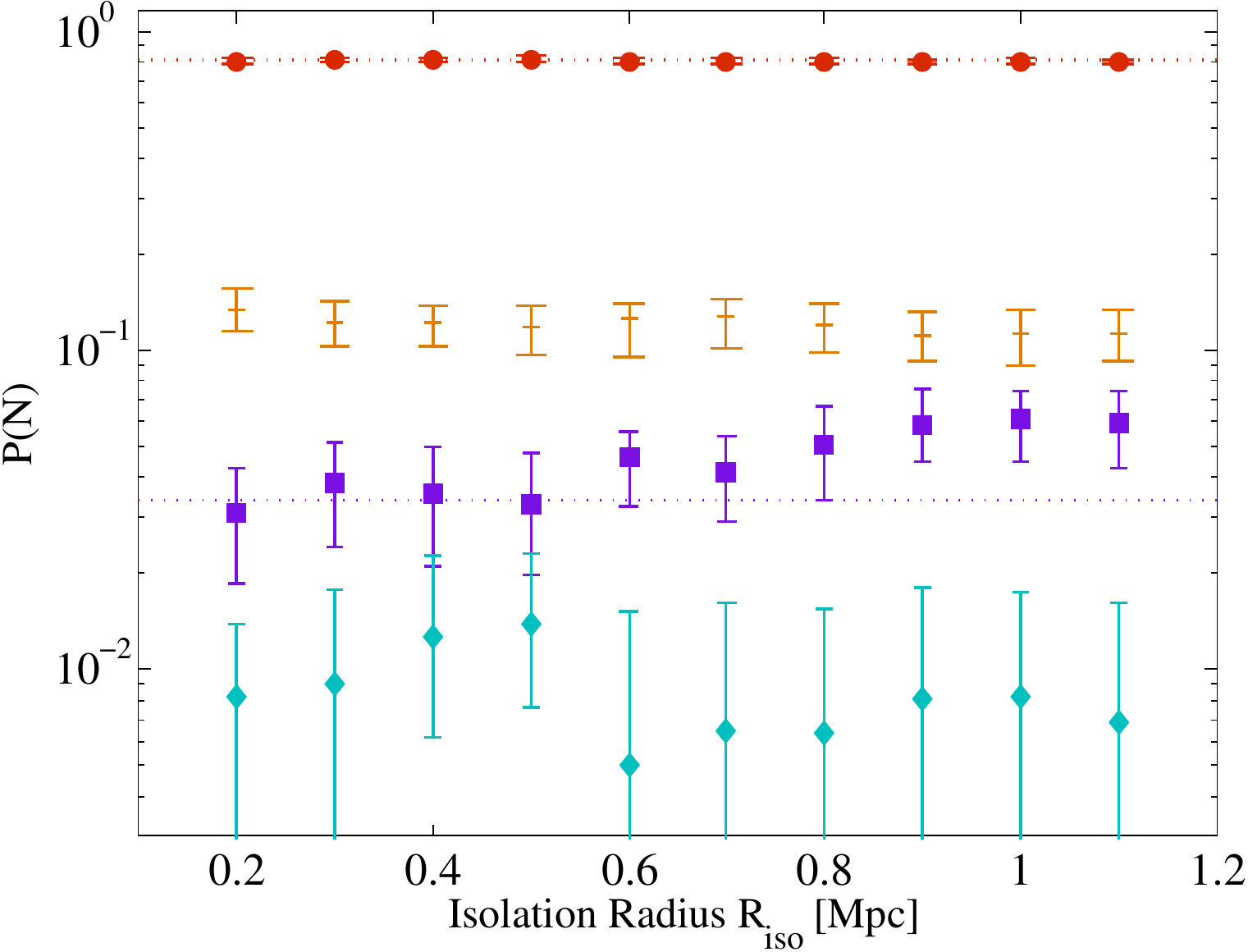}
\includegraphics[width=3.3in]{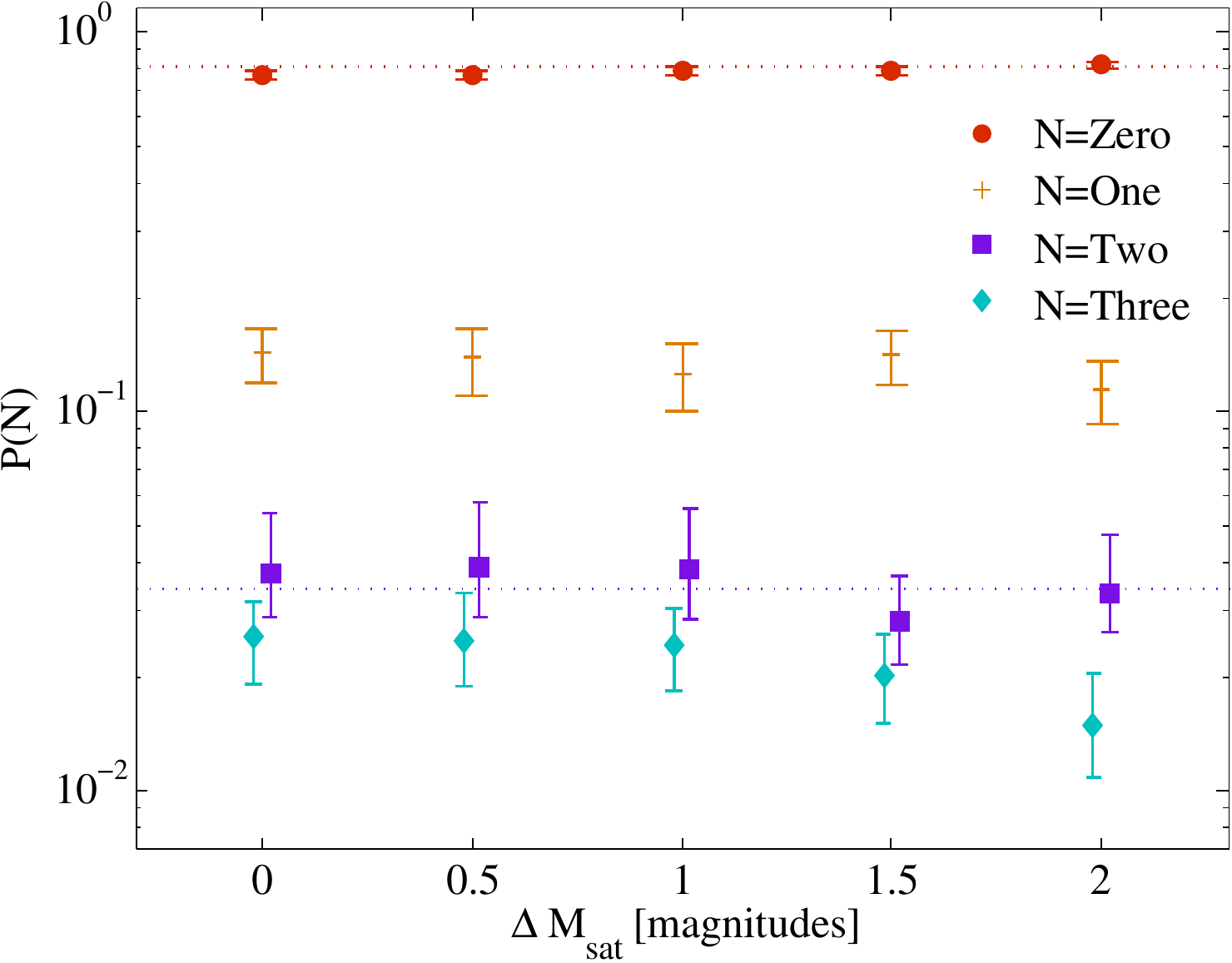}
\includegraphics[width=3.3in]{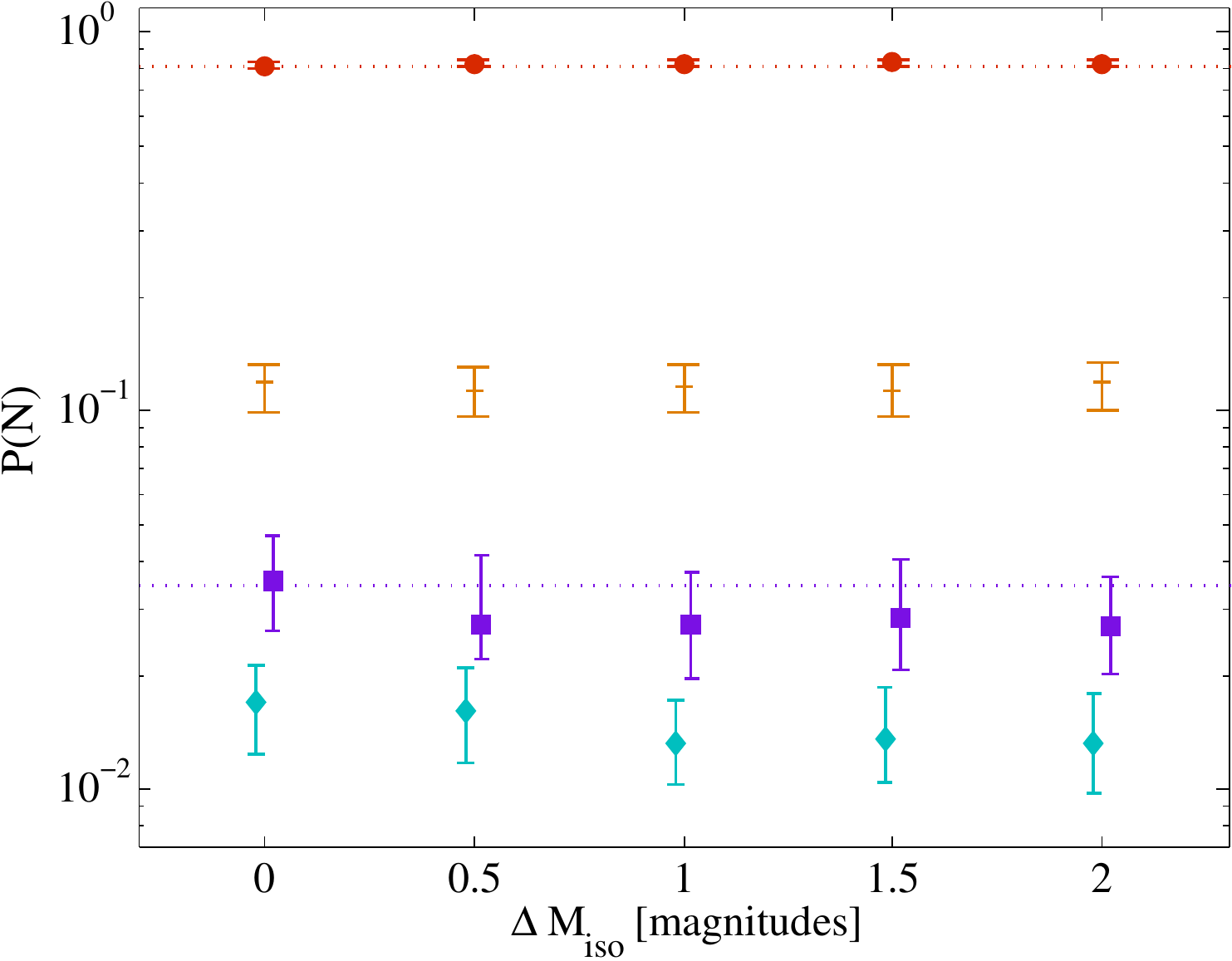}
\caption{Sensitivity of the probability of hosting $N$=0,1,2, or 3
  satellites to changes in various selection parameters. In each panel,
  one selection parameter is varied and the others are held fixed at
  our nominal values that were used in Figure~\ref{fig:results}.  Our
  results for the nominal 
  parameter values are shown as dotted lines. \emph{Top Left:}
  Dependence of probabilities on satellite search radius around the
  host galaxy.  \emph{Top Right:} Dependence of probabilities on
  variation of the isolation radius around the host galaxy,
  $R_{iso}$. \emph{Bottom Left:} The allowed magnitude disparity 
  between host and MC-like satellite, $\Delta M_{sat}$, is varied.
  Here, we search for satellites with magnitudes 2-4, 1.5-4, 1-4,
  0.5-4 and 0.1-4 magnitudes dimmer than host, plot is indexed by the
  changing minimum value.  \emph{Bottom Right:} Results with
  increasingly stringent host-neighbor relative 
  brightness limit $\Delta M_{iso}$.}
\label{fig:varyparams}
\end{figure*}

In this section we confirm the stability of our main results for the
probability of finding $N_{sat}$ MC-like satellites in a sphere of
radius $R_{sat}$ around an MW-sized host.  We vary several key
parameters defined earlier in Sections \ref{sec:MWhosts} and
\ref{sec:satellites}: $R_{sat}$ (the satellite search radius), $\Delta
M_{sat}$ (the maximum satellite magnitude relative to the host),
$R_{iso}$ (the host isolation radius), and $\Delta M_{iso}$ (the host
isolation relative magnitude limit).  The first two parameters alter
our definition of a MC-like satellite, while the latter two change
what is considered a suitable MW-like host.  We vary each of these
parameters over a reasonable range of selection criteria that might be
expected to produce an approximate analog of the MW-LMC-SMC system.
The results of this investigation are shown in
Figure~\ref{fig:varyparams}, where each parameter is varied in turn,
while holding the other parameters fixed at their nominal values.


As would naively be expected, more
satellites are detected as we increase the satellite search radius
$R_{sat}$.  However, the satellite counts are remarkably flat out to
$R_{sat} = 200$ kpc. If we very stringently require candidate MCs to lie
within $R_{sat} = 100$ kpc of their hosts (as the LMC and SMC do) then
slightly less than 3\% of MW-sized galaxies host two MC-like satellites.
On the other hand, if we expand the search radius to $200$ kpc, this
fraction becomes about 5\%.  Even expanding it to $250$ kpc (roughly
the virial radius of the MW derived in ~\citealt{Busha10c}), the
fraction of hosts with two satellites rises only to $\sim 8\%$. This
suggests that our analysis has 
largely captured the probability of true MC-analog satellites.

To further test whether we have captured the full satellite population
in our main analysis, we compute the mean $N_{sat}$ values in each of
the radial bins shown in the upper left-hand panel of the Figure.
Taking the difference between these values, and assuming a spherical
search geometry, we can then compute the number density of satellites
in bins of radius.  Because this measurement is nonnegative by
construction, we expect that stochastic noise will cause us to measure
a positive value in each bin; however, once we have measured the
satellite population as completely as is possible within the
uncertainties, the measured number density at all higher radii will be
consistent with zero.  In performing this exercise, we find that the
measured average number density rises sharply below $R_{sat} = 150$
kpc and that it is roughly flat and consistent with zero at all higher
values of $R_{sat}$.  This confirms that our fiducial value of
$R_{sat}$ captures the MC-analog population as well as is possible
within the uncertainties in our analysis.

In addition, a slight upward trend in the $N=2$ value appears at the
$1\sigma$ level as hosts become increasingly isolated from larger
neighbors (upper-right panel).  If this weak trend is real, it is most
easily explained as an effect of host formation history.  More
isolated hosts will have formed more recently, on average, so they
will have had less time to disrupt or accrete their satellites, and so
their satellite population will be enhanced relative to hosts in
denser regions.

Lastly, we note that there is very little trend with the satellite
relative-luminosity criterion $\Delta M_{sat}$ (lower left panel of
the Figure), despite the fact that we are increasing the magnitude
range considered by up to a factor of two.  Since the overall galaxy
luminosity function is not particularly steep over this magnitude
range, one might expect the satellite probability to rise
substantially when we broaden this search criterion.  However, there
is no particular reason that the luminosity function of
\emph{satellites of MW analogs} should the same as the overall
luminosity function in this range.  Our results suggest in fact that
it is not.   We may conclude from this result that satellites brighter
than the MCs are extremely rare.   


No other significant trends are observed under variation of host
parameters.  Specifically, little to no change in the results is
evident if we reject hosts with companions slightly fainter than
themselves.  This is not particularly surprising, since such galaxies
constitute only around one quarter of the sample in our primary
analysis.  Thus, we find our results to be quite robust to all
significant and reasonable parameter changes; they are not simply an
accident of the satellite search criteria we have chosen.

\subsubsection{The Stripe 82 co-added catalog}
\label{sec:stripe82}

We partially repeat the analysis for the deeper co-added data in the SDSS
equatorial stripe (Stripe 82).  The Stripe 82 catalog is not only 
deeper than the main SDSS imaging database (magnitude limit $r \approx
23.5 $) but has no spatial intersection with the Northern Galactic
Cap, offering a disjoint set of objects with which we can verify the
results. Because of the deeper photometric limit, we can consider
potential MW-like hosts a bit dimmer, near to  the completeness limit of the
main SDSS spectroscopic sample, $r=17.60$
(whereas we were limited to $r=17$, four magnitudes brighter than our
photometric limit, in the NGC). Even with this deeper magnitude cut,
there are only 1946 MW-sized galaxies in Stripe 82 that have spectra
and meet our primary isolation criteria, compared to 22,581 in the
NGC.  This sample extends to slightly higher redshift: $z=0.15$,
rather than $0.12$.

Since the statistical power of this sample is limited by its small
size, we choose to compute only one of the results for comparison to
the NGC sample.  The simplest result to compute is the PDF of
correlated galaxy counts, $p(N_{cor})$, calculated using isotropic
background estimation.  We have already shown that this result can be
systematically corrected to accurately recover $p(N_{sat})$, and there
would be no changes to this correction procedure in the Stripe 82
data set, so comparing this one result should be sufficient.  The
isotropic-background result also has the advantage of being most
directly comparable to the spectroscopic analysis we performed in
Section~\ref{sec:spectrosat} (as explained in
Section~\ref{sec:noisecount}).

All the methods described earlier apply to this analysis except for
the specifics of our choice of $z_{phot,max}$ and computation of the
loss fraction $\eta$.  A careful photo-$z$ cut is even more important
here, as the sky density of photometric objects in Stripe 82 far
exceeds that of the main SDSS imaging catalog in the north.  Here, we
use photometric redshifts computed for the full Stripe 82 co-added
catalog (Reis et al. in preparation) using the neural-network approach
of \citet{oyaizu08}.  Lacking full $p(z)$ estimates for this sample,
we compute $\eta$ from a subset of the photo-$z$ validation set
matched to the apparent magnitude and redshift distributions of our
MC-like satellites. Objects in the validation set have measured
spectroscopic redshifts but were not used to train the photo-$z$
algorithm; they are used to test the accuracy of the photo-$z$
estimates.  We can make histograms of this data set to obtain $p(z)$
distributions for different subsets and then perform an analysis
analogous to Section \ref{sec:systematics} to obtain the fractional
photo-$z$ loss, $\eta$.  We find that, for the Stripe 82 photo-$z$
values, a cut of $z_{phot,max}=0.21$ corresponds to $\eta = 0.15$,
which is acceptably small, so we use this maximum photo-$z$ cut in our
analysis.

\begin{figure}[t!]
\centering
\includegraphics[width=3in]{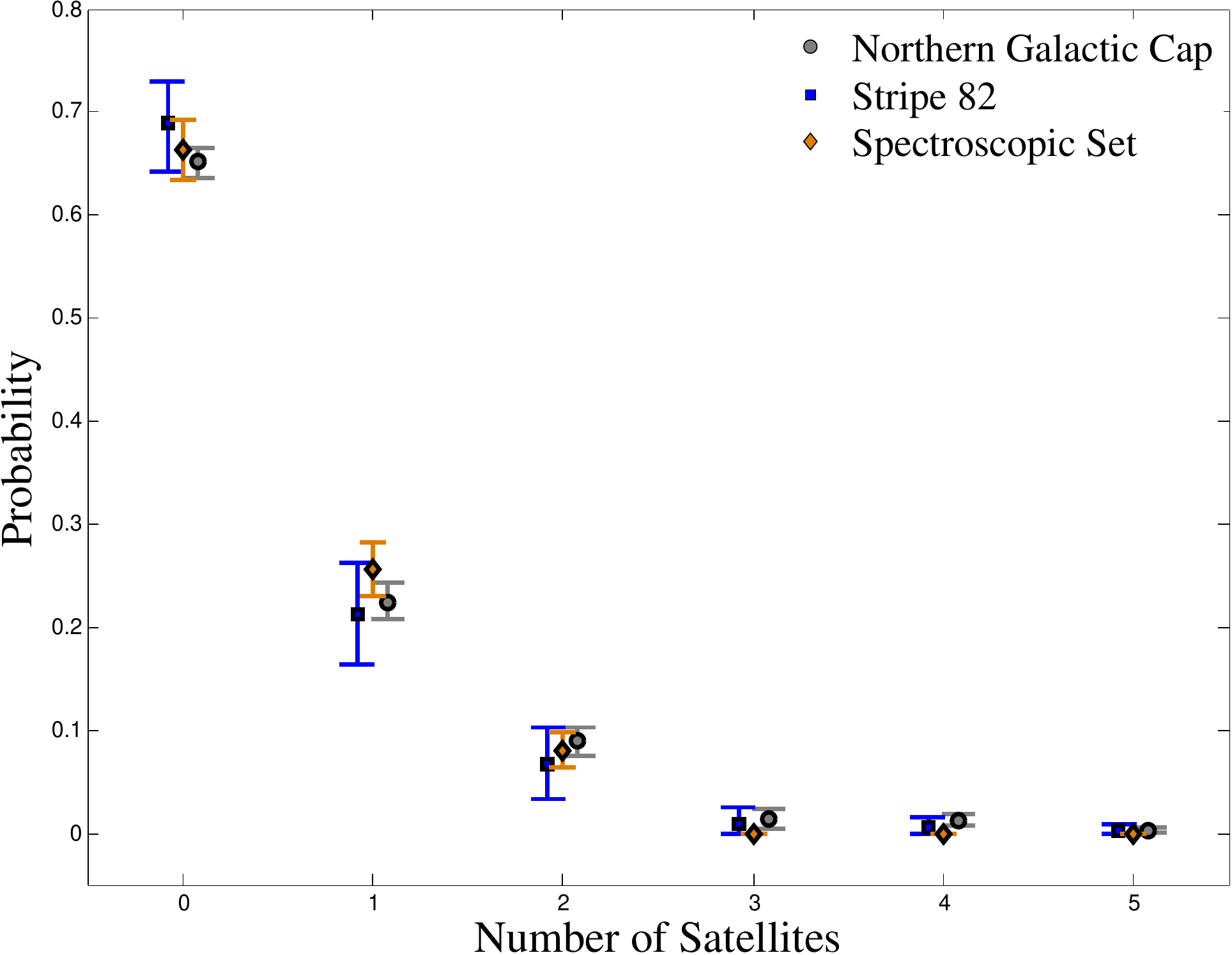}
\caption{Probability of finding $N_{cor}$ correlated objects around a
  MW analog in a cylinder with radius $R_{sat} = 150kpc$, computed
  using three different data sets.  \emph{Grey:} Results using hosts
  from the Northern Galactic Cap region of the SDSS spectroscopic main
  catalog and satellites from photometric main catalog.  \emph{Blue:}
  Results using hosts from the Stripe 82 region of the spectroscopic
  catalog and satellites from photometric Co-added data.
  \emph{Orange:} Results using hosts and satellites only from the NGC
  region of the spectroscopic main catalog.  }
\label{fig:stripe82}
\end{figure}

The $p(N_{cor})$ results obtained from Stripe 82 are in good agreement
with those obtained using the photometric catalog in the NGC in
Section~\ref{sec:primaryresults} and with the results computed
using only spectroscopic information in Section~\ref{sec:spectrosat}.
A comparison of the three $p(N_{cor})$ measurements is shown in Figure
\ref{fig:stripe82}. Since disjoint data sets yield statistically
identical results despite covering disparate ranges in redshift and
apparent magnitude, and despite having photo-$z$-induced systematic
errors computed with different algorithms, we can be confident that
our results do not depend strongly on these details.

\section{Conclusions}
\label{sec:conclusions}

We have investigated the occurrence of dwarf satellites with
luminosities similar to the Magellanic Clouds around host galaxies
with environment and luminosity similar to the MW.  Our analysis uses
spectroscopic data from SDSS to identify isolated MW-like galaxies and
then searches the SDSS photometric data for potential satellites
between two and four magnitudes fainter than these hosts.  The primary
result, summarized in Table~\ref{tab:results-sphere} is the
probability distribution of hosting $N_{sat}$ satellites similar to
the LMC and SMC.  We find that, of our 22,581 MW-luminosity host
galaxies, $81\%$ have zero satellites as bright as the Magellanic
Clouds, $11\%$ have one such satellite, and only $3.5\%$ host two such
satellites.

The main source of uncertainty in our analysis is the presence of
projected foreground and background objects along the line of sight to
each MW analog. We correct for these in two stages.  First, we reject
most background objects with a rough photometric-redshift cut, and
then we statistically correct for the remaining background objects by
comparing the counts around MW analogs to counts in areas of the sky
that do not contain MW-like objects.  Because the background-noise
level (rather than the sample size) is the dominant source of
statistical uncertainty in our main analysis, the best potential for
improving upon the precision of our results would come from an 
improvement in the photometric redshift estimates of faint objects,
which would allow a more stringent initial background rejection.

The specific manner in which the
background-estimation fields are selected has important implications for our
final results.  Fields chosen at random positions on the
sky do not account for the clustering of galaxies, which will enhance
line-of-sight projections around our chosen hosts, although an
approximate correction can be calculated. We emphasize that such a
correction is needed even when perfect spectroscopic information is
available.  Alternatively, the background fields may be chosen as
annuli around each host, outside of the satellite search radius.  This
accounts for both random and correlated line-of-sight projections, and
it is the technique we use for our main results.  However, the optimum
radius for these annuli is not entirely obvious, and this introduces a
small additional systematic uncertainty into our results.  Allowing
for this systematic error, it is possible that the percentage of MW
analogs that host two MC-analog satellites could be as high as $\sim
5\%$ (see Table~\ref{tab:results-sphere}).

Nevertheless, the clear qualitative conclusion is that the presence of
the LMC and SMC makes the MW quite unusual among the population of
galaxies with similar luminosity.  This is broadly in agreement with
earlier observational studies that found $\la 1$ satellite on average
around typical bright galaxies \citep{zaritsky93, lorrimer94,
  chen_etal:06, james10}.  In fact, when we compute the average number
of satellites from our measured $p(N_{sat})$ distribution in
Table~\ref{tab:results-sphere}, we find $\langle N_{sat} \rangle =
0.3$, which is \emph{lower} than the mean values reported in those
studies.  However, because previous authors considered a much wider
range of host and satellite luminosities than we consider here, we do
not expect more than qualitative agreement in any case.  Similarly, we
do not find a statistically significant difference in $p(N_{sat})$ for
red versus blue galaxies, and this appears to be in conflict with the
results of \citet{lorrimer94} and \citet{chen:08}, who find that
early-type galaxies host significantly more galaxies than late-types
do.  However, the broad range of host luminosities considered in those
earlier studies, combined with the different luminosity distributions of
early and late types, means that the underlying trend they found could
be with luminosity, rather than color.

Our results are useful for understanding the larger cosmological
context of the Milky Way Galaxy.  There is a striking agreement
between our results and the predictions of recent high-resolution
cosmological N-body simulations.  For example, \citep{MBK10} found
that MW-sized dark-matter halos hosted two MC-sized subhalos only $\la
10\%$ of the time in the Millennium II simulation \cite{MillenniumII}.
Similar results are obtained with the Bolshoi simulation; the
consistency between our results and the Bolshoi predictions will be
discussed in detail by \cite{Busha10b}.  This agreement
constitutes an important confirmation of the cold dark matter paradigm
for galaxy formation.

Our result also indicates that the MW is somewhat unusual among
galaxies of similar luminosity at least in terms of its satellite
population.  A major philosophical underpinning of our cosmology is
the Copernican principle, which holds that we do not observe the
Universe from any privileged position, except insofar as such a
position is required for our existence (\emph{e.g.}, our very atypical
position on a rocky planet with an atmosphere).  Since there is no
obvious anthropic requirement on the number of bright satellite
companions to the MW, it would not be unreasonable to wonder whether
our results present a challenge to the Copernican principle.

Applied to the expected properties of our home galaxy, a reasonable
statement of the principle is that the Milky Way should be consistent
with a galaxy chosen at random from the stellar-mass-weighted galaxy
population at large.  It is important to note that this does
\emph{not} necessarily mean that the MW should be ``typical'' in all
possible respects.  In particular, it is not particularly unexpected
that a randomly selected galaxy will be a $\sim 2\sigma$ outlier by at
least one measure.  Moreover, there is now reasonably strong evidence
that the LMC and SMC were recently accreted by the MW and are on their
first pass through the halo \citep[\eg,][]{besla07, Busha10c}.  If
this is true, then the presence of the Magellanic Clouds may be a
transient event in the formation history of the MW, implying that the
MW is not fundamentally unusual in any way.  Thus, we may conclude
that the unusually large population of bright satellites around the MW
can likely be ascribed to happenstance and presents no special
challenge to our basic cosmological paradigm.

\acknowledgments 
This project has made use of data from the Sloan Digital Sky Survey
(SDSS).  Funding for the SDSS has been provided by the Alfred P. Sloan
Foundation, the Participating Institutions, the National Aeronautics
and Space Administration, the National Science Foundation, the
U.S. Department of Energy, the Japanese Monbukagakusho, and the Max
Planck Society. The SDSS Web site is http://www.sdss.org/.

The SDSS is managed by the Astrophysical Research Consortium (ARC) for
the Participating Institutions. The Participating Institutions are The
University of Chicago, Fermilab, the Institute for Advanced Study, the
Japan Participation Group, The Johns Hopkins University, Los Alamos
National Laboratory, the Max-Planck-Institute for Astronomy (MPIA),
the Max-Planck-Institute for Astrophysics (MPA), New Mexico State
University, University of Pittsburgh, Princeton University, the United
States Naval Observatory, and the University of Washington.

We thank Marcelle Soares-Santos, Ribamar Reis, Huan Lin, and Jim Annis
for providing us with the SDSS Stripe 82 co-add data and photometric
redshifts in a usable form, as well as for providing the spectroscopic
validation set for these data.  We thank Carlos Cunha for several
useful discussions about photometric redshifts, and for providing
photo-$z$s for the SDSS main sample.  We thank Sarah Hansen for useful
discussion about statistical background correction, Jeff Newman for
helpful advice about correlation functions, and Marla Geha and
Louie Strigari for useful input about the satellites of the Milky
Way.  The purity and completeness of our sample was tested using halos
from the Bolshoi simulation.  We thank Anatoly Klypin and Joel Primack
for providing access to this simulation, which was using NASA Advanced
Supercomputing resources at NASA Ames Research Center.

RHW, LL, and MTB were supported by the National Science Foundation
under grant NSF AST-0807312; RHW, BFG and PSB received support from
the U.S. Department of Energy under contract number DE-AC02-76SF00515.

\bibliographystyle{apj}
\bibliography{master}

\begin{thebibliography}{38}
\expandafter\ifx\csname natexlab\endcsname\relax\def\natexlab#1{#1}\fi

\bibitem[{{Abazajian} {et~al.}(2009)}]{Abazajian09}
{Abazajian}, K.~N., {et~al.} 2009, \apjs, 182, 543

\bibitem[{{Belokurov} {et~al.}(2007){Belokurov}, {Zucker}, {Evans}, {Kleyna},
  {Koposov}, {Hodgkin}, {Irwin}, {Gilmore}, {Wilkinson}, {Fellhauer},
  {Bramich}, {Hewett}, {Vidrih}, {De Jong}, {Smith}, {Rix}, {Bell}, {Wyse},
  {Newberg}, {Mayeur}, {Yanny}, {Rockosi}, {Gnedin}, {Schneider}, {Beers},
  {Barentine}, {Brewington}, {Brinkmann}, {Harvanek}, {Kleinman}, {Krzesinski},
  {Long}, {Nitta}, \& {Snedden}}]{Belokurov07b}
{Belokurov}, V., {et~al.} 2007, \apj, 654, 897

\bibitem[{{Besla} {et~al.}(2007){Besla}, {Kallivayalil}, {Hernquist},
  {Robertson}, {Cox}, {van der Marel}, \& {Alcock}}]{besla07}
{Besla}, G., {Kallivayalil}, N., {Hernquist}, L., {Robertson}, B., {Cox},
  T.~J., {van der Marel}, R.~P., \& {Alcock}, C. 2007, \apj, 668, 949

\bibitem[{{Blanton} {et~al.}(2005{\natexlab{a}}){Blanton}, {Lupton},
  {Schlegel}, {Strauss}, {Brinkmann}, {Fukugita}, \& {Loveday}}]{Blanton05b}
{Blanton}, M.~R., {Lupton}, R.~H., {Schlegel}, D.~J., {Strauss}, M.~A.,
  {Brinkmann}, J., {Fukugita}, M., \& {Loveday}, J. 2005{\natexlab{a}}, \apj,
  631, 208

\bibitem[{{Blanton} \& {Roweis}(2007)}]{BlantonRoweis}
{Blanton}, M.~R., \& {Roweis}, S. 2007, \aj, 133, 734

\bibitem[{{Blanton} {et~al.}(2005{\natexlab{b}}){Blanton}, {Schlegel},
  {Strauss}, {Brinkmann}, {Finkbeiner}, {Fukugita}, {Gunn}, {Hogg},
  {Ivezi{\'c}}, {Knapp}, {Lupton}, {Munn}, {Schneider}, {Tegmark}, \&
  {Zehavi}}]{Blanton05}
{Blanton}, M.~R., {et~al.} 2005{\natexlab{b}}, \aj, 129, 2562

\bibitem[{{Boylan-Kolchin} {et~al.}(2010){Boylan-Kolchin}, {Springel}, {White},
  \& {Jenkins}}]{MBK10}
{Boylan-Kolchin}, M., {Springel}, V., {White}, S.~D.~M., \& {Jenkins}, A. 2010,
  \mnras, 406, 896

\bibitem[{{Boylan-Kolchin} {et~al.}(2009){Boylan-Kolchin}, {Springel}, {White},
  {Jenkins}, \& {Lemson}}]{MillenniumII}
{Boylan-Kolchin}, M., {Springel}, V., {White}, S.~D.~M., {Jenkins}, A., \&
  {Lemson}, G. 2009, \mnras, 398, 1150

\bibitem[{{Bullock}(2010)}]{bullock10}
{Bullock}, J.~S. 2010, ArXiv e-prints, arXiv:1009.4505 [astro-ph.CO]

\bibitem[{Busha {et~al.}(2010{\natexlab{a}})Busha, Marshall, Wechsler, \&
  Klypin}]{Busha10c}
Busha, M., Marshall, P., Wechsler, R.~H., \& Klypin, A. 2010{\natexlab{a}},
  ApJ, submitted

\bibitem[{Busha {et~al.}(2010{\natexlab{b}})Busha, Wechsler, Behroozi,
  {et~al.}}]{Busha10b}
Busha, M., Wechsler, R.~H., Behroozi, P., {et~al.} 2010{\natexlab{b}}, in
  preparation

\bibitem[{{Busha} {et~al.}(2010){Busha}, {Alvarez}, {Wechsler}, {Abel}, \&
  {Strigari}}]{Busha10a}
{Busha}, M.~T., {Alvarez}, M.~A., {Wechsler}, R.~H., {Abel}, T., \& {Strigari},
  L.~E. 2010, \apj, 710, 408

\bibitem[{{Chen}(2008)}]{chen:08}
{Chen}, J. 2008, \aap, 484, 347

\bibitem[{{Chen} {et~al.}(2006){Chen}, {Kravtsov}, {Prada}, {Sheldon},
  {Klypin}, {Blanton}, {Brinkmann}, \& {Thakar}}]{chen_etal:06}
{Chen}, J., {Kravtsov}, A.~V., {Prada}, F., {Sheldon}, E.~S., {Klypin}, A.~A.,
  {Blanton}, M.~R., {Brinkmann}, J., \& {Thakar}, A.~R. 2006, \apj, 647, 86

\bibitem[{{Conroy} {et~al.}(2005){Conroy}, {Newman}, {Davis}, {Coil}, {Yan},
  {Cooper}, {Gerke}, {Faber}, \& {Koo}}]{conroy05}
{Conroy}, C., {et~al.} 2005, \apj, 635, 982

\bibitem[{{Conroy} {et~al.}(2007){Conroy}, {Prada}, {Newman}, {Croton}, {Coil},
  {Conselice}, {Cooper}, {Davis}, {Faber}, {Gerke}, {Guhathakurta}, {Klypin},
  {Koo}, \& {Yan}}]{conroy07a}
---. 2007, \apj, 654, 153

\bibitem[{{Cunha} {et~al.}(2009){Cunha}, {Lima}, {Oyaizu}, {Frieman}, \&
  {Lin}}]{cunha}
{Cunha}, C.~E., {Lima}, M., {Oyaizu}, H., {Frieman}, J., \& {Lin}, H. 2009,
  \mnras, 396, 2379

\bibitem[{{Davis} \& {Peebles}(1983)}]{DavisPeebles83}
{Davis}, M., \& {Peebles}, P.~J.~E. 1983, \apj, 267, 465

\bibitem[{Holmberg(1969)}]{holmberg69}
Holmberg, E. 1969, Arkiv f\"{o}r Astronomi, 5, 305

\bibitem[{{James} \& {Ivory}(2010)}]{james10}
{James}, P.~A., \& {Ivory}, C.~F. 2010, ArXiv e-prints, arXiv:1009.2875
  [astro-ph.CO]

\bibitem[{{Klypin} {et~al.}(1999{\natexlab{a}}){Klypin}, {Gottl{\"o}ber},
  {Kravtsov}, \& {Khokhlov}}]{Klypin99b}
{Klypin}, A., {Gottl{\"o}ber}, S., {Kravtsov}, A.~V., \& {Khokhlov}, A.~M.
  1999{\natexlab{a}}, \apj, 516, 530

\bibitem[{{Klypin} {et~al.}(1999{\natexlab{b}}){Klypin}, {Kravtsov},
  {Valenzuela}, \& {Prada}}]{Klypin99}
{Klypin}, A., {Kravtsov}, A.~V., {Valenzuela}, O., \& {Prada}, F.
  1999{\natexlab{b}}, \apj, 522, 82

\bibitem[{{Klypin} {et~al.}(2010){Klypin}, {Trujillo-Gomez}, \&
  {Primack}}]{Bolshoi}
{Klypin}, A., {Trujillo-Gomez}, S., \& {Primack}, J. 2010, ArXiv e-prints,
  arXiv:1002.3660 [astro-ph:CO]

\bibitem[{{Koposov} {et~al.}(2008){Koposov}, {Belokurov}, {Evans}, {Hewett},
  {Irwin}, {Gilmore}, {Zucker}, {Rix}, {Fellhauer}, {Bell}, \&
  {Glushkova}}]{Koposov08}
{Koposov}, S., {et~al.} 2008, \apj, 686, 279

\bibitem[{{Kravtsov}(2010)}]{kravtsov10}
{Kravtsov}, A. 2010, Advances in Astronomy, 2010

\bibitem[{{Lorrimer} {et~al.}(1994){Lorrimer}, {Frenk}, {Smith}, {White}, \&
  {Zaritsky}}]{lorrimer94}
{Lorrimer}, S.~J., {Frenk}, C.~S., {Smith}, R.~M., {White}, S.~D.~M., \&
  {Zaritsky}, D. 1994, \mnras, 269, 696

\bibitem[{{Madau} {et~al.}(2008){Madau}, {Diemand}, \& {Kuhlen}}]{Madau08}
{Madau}, P., {Diemand}, J., \& {Kuhlen}, M. 2008, \apj, 679, 1260

\bibitem[{{Moore} {et~al.}(1999){Moore}, {Ghigna}, {Governato}, {Lake},
  {Quinn}, {Stadel}, \& {Tozzi}}]{Moore99b}
{Moore}, B., {Ghigna}, S., {Governato}, F., {Lake}, G., {Quinn}, T., {Stadel},
  J., \& {Tozzi}, P. 1999, \apjl, 524, L19

\bibitem[{{Oyaizu} {et~al.}(2008){Oyaizu}, {Lima}, {Cunha}, {Lin}, {Frieman},
  \& {Sheldon}}]{oyaizu08}
{Oyaizu}, H., {Lima}, M., {Cunha}, C.~E., {Lin}, H., {Frieman}, J., \&
  {Sheldon}, E.~S. 2008, \apj, 674, 768

\bibitem[{{Prada} {et~al.}(2003){Prada}, {Vitvitska}, {Klypin}, {Holtzman},
  {Schlegel}, {Grebel}, {Rix}, {Brinkmann}, {McKay}, \&
  {Csabai}}]{prada_etal:03}
{Prada}, F., {et~al.} 2003, \apj, 598, 260

\bibitem[{{Strateva} {et~al.}(2001){Strateva}, {Ivezi{\'c}}, {Knapp},
  {Narayanan}, {Strauss}, {Gunn}, {Lupton}, {Schlegel}, {Bahcall}, {Brinkmann},
  {Brunner}, {Budav{\'a}ri}, {Csabai}, {Castander}, {Doi}, {Fukugita}, {Gy{\H
  o}ry}, {Hamabe}, {Hennessy}, {Ichikawa}, {Kunszt}, {Lamb}, {McKay},
  {Okamura}, {Racusin}, {Sekiguchi}, {Schneider}, {Shimasaku}, \&
  {York}}]{Strateva01}
{Strateva}, I., {et~al.} 2001, \aj, 122, 1861

\bibitem[{{Tollerud} {et~al.}(2008){Tollerud}, {Bullock}, {Strigari}, \&
  {Willman}}]{Tollerud08}
{Tollerud}, E.~J., {Bullock}, J.~S., {Strigari}, L.~E., \& {Willman}, B. 2008,
  \apj, 688, 277

\bibitem[{{van den Bergh}(2000)}]{vandenBergh}
{van den Bergh}, S. 2000, \pasp, 112, 529

\bibitem[{{Walsh} {et~al.}(2009){Walsh}, {Willman}, \& {Jerjen}}]{Walsh09}
{Walsh}, S.~M., {Willman}, B., \& {Jerjen}, H. 2009, \aj, 137, 450

\bibitem[{{Wechsler} {et~al.}(2006){Wechsler}, {Zentner}, {Bullock},
  {Kravtsov}, \& {Allgood}}]{Wechsler06}
{Wechsler}, R.~H., {Zentner}, A.~R., {Bullock}, J.~S., {Kravtsov}, A.~V., \&
  {Allgood}, B. 2006, \apj, 652, 71

\bibitem[{{York} {et~al.}(2000)}]{york_etal:00}
{York}, D.~G., {et~al.} 2000, \aj, 120, 1579

\bibitem[{{Zaritsky} {et~al.}(1993){Zaritsky}, {Smith}, {Frenk}, \&
  {White}}]{zaritsky93}
{Zaritsky}, D., {Smith}, R., {Frenk}, C., \& {White}, S.~D.~M. 1993, \apj, 405,
  464

\bibitem[{{Zehavi} {et~al.}(2010){Zehavi}, {Zheng}, {Weinberg}, {Blanton},
  {Bahcall}, {Berlind}, {Brinkmann}, {Frieman}, {Gunn}, {Lupton}, {Nichol},
  {Percival}, {Schneider}, {Skibba}, {Strauss}, {Tegmark}, \&
  {York}}]{zehavi_etal:10}
{Zehavi}, I., {et~al.} 2010, ArXiv e-prints, arXiv:1005.2413 [astro-ph.CO]

\end{thebibliography}
\end{document}